\documentclass[5p,times,twocolumn]{elsarticle}

\usepackage{lineno,hyperref}
\usepackage{import}
\modulolinenumbers[5]

\journal{Future Generation Computer Systems}

\usepackage{color,soul}

\usepackage{todonotes}
\usepackage{gensymb}
\usepackage{caption}
\usepackage{csvsimple}
\usepackage{booktabs, tabularx, wrapfig}
\usepackage{array, ltablex, multirow}
\usepackage{placeins}
\usepackage{graphicx}
\usepackage{subcaption}

\usepackage[nonumberlist,nogroupskip]{glossaries}
\usepackage{xspace}
\usepackage[binary-units=true,per-mode=symbol]{siunitx}
\usepackage{blindtext,rotating}
\usepackage{tablefootnote}
\usepackage{threeparttable}

\setglossarystyle{super}

\hypersetup{%
  colorlinks=false,
  linkbordercolor=[rgb]{0.6,0.6,0.6},
  pdfborderstyle={/S/U/W 1}
}

\newacronym{ti}{$T_{i}$}{indoor air temperature, $^{\circ}$C}
\newacronym{tout}{$T_{out}$}{outdoor air temperature, $^{\circ}$C}
\newacronym{top}{$T_{o}$}{operative air temperature}
\newacronym{ts}{$T_{s}$}{supply air temperature, $^{\circ}$C}
\newacronym{v}{$\dot{V}$}{measured volumetric flow rate, L/s}
\newacronym{vsp}{$\dot{V}_{sp}$}{volumetric flow rate set-point, L/s}
\newacronym{tsph}{$T_{sp,h}$}{heating zone temperature set-point, $^{\circ}$C}
\newacronym{tspc}{$T_{sp,c}$}{cooling zone temperature set-point}
\newacronym{ghi}{$G_{hi}$}{global horizontal irradiance, W/m$^2$}

\newacronym{iot}{IoT}{Internet of Things}
\newacronym{dl}{DL}{Deep Learning}
\newacronym{ml}{ML}{Machine Learning}
\newacronym{dnn}{DNN}{Deep Neural Network}
\newacronym{nn}{NN}{Neural Network}
\newacronym{soc}{SoC}{System on-Chip}
\newacronym{fp}{FP}{Floating-Point}
\newacronym{gemm}{GEMM}{General Matrix-Matrix Multiplication}
\newacronym{gemmop}{GEMM-Op}{General Matrix-Matrix Operation}
\newacronym{gpu}{GPU}{Graphic Processing Unit}
\newacronym{pulp}{PULP}{Parallel Ultra-Low-Power}
\newacronym{ce}{CE}{Computing Element}
\newacronym{hpc}{HPC}{High Performance Computing}
\newacronym{fma}{FMA}{Fused Multiply-Add}
\newacronym{fpu}{FPU}{Floating-Point Unit}
\newacronym{cnn}{CNN}{Convolutional Neural Network}
\newacronym{soa}{SoA}{State of the Art}
\newacronym{tcdm}{TCDM}{Tightly-Coupled Data Memory}
\newacronym{dmac}{DMAC}{Direct Memory Access Controller}
\newacronym{hwpe}{HWPE}{Hardware Processing Engine}
\newacronym{hci}{HCI}{Heterogeneous Cluster Interconnect}
\newacronym{fncomp}{FNCOMP}{Floating-Point Non-Computational Operations}
\newacronym{bfs}{BFS}{Breadth-First Search}

\newcommand{\riscv}{RISC-V\xspace}
\newcommand{\etal}{\textit{et al.}\xspace}
\newcommand{\dbit}[1]{{#1}{-bit}}

\newcommand{\simdsquare}{SIMD\textsuperscript{2}\xspace}
\newcommand{\redmuleHalf}{RedMulE\textsubscript{12x4}\xspace}
\newcommand{\redmuleByte}{RedMulE\textsubscript{12x8}\xspace}

\begin{document}

\begin{frontmatter}

\title{RedMule: A Mixed-Precision Matrix-Matrix Operation Engine for Flexible and Energy-Efficient On-Chip Linear Algebra and TinyML Training Acceleration}

\author[unibo]{Yvan Tortorella}
\ead{yvan.tortorella@unibo.it}
\author[eth]{Luca Bertaccini}
\ead{lbertaccini@iis.ee.ethz.ch}
\author[unibo,eth]{Luca Benini}
\ead{lbenini@iis.ee.ethz.ch}
\author[unibo]{Davide Rossi}
\ead{davide.rossi@unibo.it}
\author[unibo]{Francesco Conti}
\ead{f.conti@unibo.it}


\address[unibo]{University of Bologna, Bologna, Italy}
\address[eth]{ETH Zurich, Zurich, Switzerland}

\begin{abstract}
The increasing interest in TinyML, i.e., near-sensor machine learning on power budgets of a few tens of mW, is currently pushing toward enabling TinyML-class training as opposed to inference only. Current training algorithms, based on various forms of error and gradient backpropagation, rely on floating-point matrix operations to meet the precision and dynamic range requirements. So far, the energy and power cost of these operations has been considered too high for TinyML scenarios. This paper addresses the open challenge of near-sensor training on a few mW power budget and presents RedMulE - Reduced-Precision Matrix Multiplication Engine, a low-power specialized accelerator conceived for multi-precision floating-point General Matrix-Matrix Operations (GEMM-Ops) acceleration, supporting FP16, as well as hybrid FP8 formats, with $\{sign, exponent,mantissa\}=(\{1,4,3\}, \{1,5,2\})$.
We integrate RedMule into a Parallel Ultra-Low-Power (PULP) cluster containing eight energy-efficient RISC-V cores sharing a tightly-coupled data memory and implement the resulting system in a 22 nm technology. At its best efficiency point (@ 470 MHz, 0.65 V), the RedMulE-augmented PULP cluster achieves 755 GFLOPS/W and 920 GFLOPS/W during regular General Matrix-Matrix Multiplication (GEMM), and up to 1.19 TFLOPS/W and 1.67 TFLOPS/W when executing GEMM-Ops, respectively, for FP16 and FP8 input/output tensors. In its best performance point (@ 613 MHz, 0.8 V), RedMulE achieves up to 58.5 GFLOPS and 117 GFLOPS for FP16 and FP8, respectively, with 99.4\% utilization of the array of Computing Elements and consuming less than 60 mW on average, thus enabling on-device training of deep learning models in TinyML application scenarios while retaining the flexibility
to tackle other classes of common linear algebra problems efficiently.
\end{abstract}


\end{frontmatter}

\section{Introduction}\label{section:introduction}

In the last few years, the number of \gls{iot} devices connected and executing \gls{ml} and, in particular, \gls{dl} based algorithms such as \glspl{dnn} increased considerably. To reduce the amount of data sent over the network, improve energy efficiency, and prevent network congestion, the computation has been moved increasingly from data centers to energy-efficient \gls{iot} end-nodes with low power budgets (a few mW average, a hundred mW peak)~\cite{edge_computing}, giving rise to the Tiny-ML field of research and application.

Extreme-edge applications like training and inference of \glspl{nn}, graph analysis and manipulation~\cite{j_gilbert1, j_gilbert2}, short-distance problems~\cite{semiring}, and model-based control rely on \glspl{gemm} or \glspl{gemmop} as the most significant kernel. \glspl{gemmop} are operations that share the same structure of a \gls{gemm} but replace the canonical multiply/add with other mapping and reduction operations~\cite{generalized_mm}. Due to the similarity of these computational patterns, it has recently been proposed~\cite{simd2} to augment TensorCores with \glspl{gemmop} support, thereby extending their acceleration capabilities to a broader class of applications. There is not yet an equal contribution targeting ultra-low-power embedded systems.

In desktop, mobile, and data center computing, single and double-precision \gls{fp} operations are typically employed for \gls{dl} and linear algebra applications, providing high accuracy at an acceptable area and energy cost. However, on embedded devices, power and area constraints are much tighter. Recently, a significant effort has gone into adapting linear algebra-based algorithms as well as online learning~\cite{cont_learn} to low-precision formats, such as \gls{fp}16~\cite{nvidia_mixp, intel_mixp} and \gls{fp}8~\cite{fp8_train, hyb_fp8}, while incurring in little accuracy loss. These algorithmic advancements enabled performance and energy efficiency gains~\cite{trans_fp, minifloat_nn}, opening the way for deploying continual learning and adaptation of \gls{dl} models on extreme-edge computing systems such as smart wearable devices. However, the computational capabilities of microcontroller units (MCUs), typically used in these devices, are minimal, especially concerning the execution of \gls{fp} operations.

In this paper, we present RedMulE (Reduced-precision matrix Multiplication Engine), the first TinyML-class open-source hardware accelerator supporting mixed \gls{fp} precision (Hybrid-\gls{fp}8 and \gls{fp}16) linear algebra that is tightly integrated within \riscv-based \gls{pulp} clusters. RedMule is highly parametric, supporting different configurations tunable at design time. It shares an L1 memory with the cluster cores, thus allowing fine-grained cooperation between the hardware accelerator and the cores. RedMulE accelerates low-precision \glspl{gemm} (the key kernels behind \gls{nn} training algorithms), enabling on-chip learning on Tiny-ML-class ultra-low-power \gls{soc}. Furthermore, it supports a broader family of operations called \glspl{gemmop}~\cite{simd2}, thus targeting an extensive set of applications.

We prototyped our design within an 8-core \gls{pulp} cluster in \SI{22}{nm} CMOS technology, instantiating a RedMulE instance with 48 internal \glspl{ce}. RedMulE occupies only \SI{0.15}{\milli\meter\squared}, accounting for $24\%$ of the entire cluster area. It achieves up to $15\times$ speedup during regular \gls{fp}16 \glspl{gemm} and up to $62\times$ during \glspl{gemmop} compared to parallel execution on the \riscv cores, reaching up to \SI{58.5}{GFLOPS} ($99.4\%$ \glspl{ce} utilization) at \SI{613}{MHz} and \SI{0.8}{V}. In its best efficiency point, i.e. \SI{470}{MHz} at \SI{0.65}{V}, RedMulE achieves up to \SI{772}{GFLOPS/W} and \SI{1.19}{TFLOPS/W} energy efficiency for \gls{gemm} and \glspl{gemmop} respectively, while reaching \SI{44.8}{GFLOPS}. When computing on \gls{fp}8 input/output tensors representation, a 96 \glspl{ce} RedMulE implementation reaches up to \SI{117}{GFLOPS} at \SI{613}{MHz} and \SI{0.8}{V}, achieving up to \SI{920}{GFLOPS/W}, an energy efficiency comparable to INT8 inference accelerators.

\section{Related Work}\label{section:related}

The strong interest in executing linear algebra-based algorithms like inference and training of \glspl{nn} led to the development of various hardware platforms specialized in this task, spanning from data-centers computing systems to ultra-low-power embedded platforms~\cite{9926331}. NVIDIA's recent Hopper H100~\cite{h_100} \gls{gpu} is the most representative example of data-center computing platform for \gls{dl} tasks like inference and training of \glspl{nn}. The H100 achieves \SI{1978}{TFLOPS} at \SI{700}{\watt} power consumption and can be used to train huge \gls{nn} models like transformers by using narrow \gls{fp}8 formats.

On the other hand, enabling the execution of \gls{dl}-based algorithms on ultra-low-power TinyML \glspl{soc} for extreme-edge devices such as smart wearable systems is challenging due to the strict power, energy, and cost constraints imposed. Extreme-edge inference is achievable in practical cases since it can be performed employing low-precision integer arithmetic, which reduces the model's memory footprint and increases the energy efficiency of the underlying architecture with a limited accuracy loss~\cite{edge_intell, 9627710}. On the contrary, extreme-edge \glspl{nn} training faces large memory requirements and the need for \gls{fp} calculations, which typically leads to power envelopes exceeding the TinyML constraints~\cite{9627710, JIN2021908}. In this section, we focus on embedded platforms emphasizing edge training at moderate power.

\subsection{Inference Accelerators}
Hardware accelerators specialized for low-power \gls{dl} inference provide attractive alternatives to software-based executions~\cite{9627710, ml_survey2}. Diana~\cite{diana}, a low-power \gls{nn} \gls{soc}, features a digital \gls{nn} inference accelerator and an analog in-memory-computing core integrated within a shared memory subsystem working only with narrow integer formats. DNPU~\cite{dnpu} is a fully-digital energy-efficient \gls{dl} processor for convolutional and recursive \gls{nn} inference acceleration designed in \SI{65}{nm} technology and based on a heterogeneous architecture supporting \dbit{16} fixed-point arithmetic. Gemmini~\cite{gemmini} is a $16\times16$ systolic accelerator designed for inference of deep \glspl{nn} with \dbit{8} multiply-accumulate units with runtime-programmable weight stationary and output stationary dataflows.

\subsection{On-Device Learning}
On-device learning is an emerging and open challenge concerning training \gls{dl} models on ultra-low-power general-purpose microcontrollers. To reach this aim, many works investigated algorithms like direct feedback alignment or equilibrium propagation. However, such methods have been demonstrated to be less effective than the classical backpropagation method due to severe convergence difficulties~\cite{cfrenkel}. TinyOL~\cite{tinyol} and~\cite{ramanathan} focus on training \glspl{nn} using the low-budget Arduino Nano microcontroller based on Cortex-M core. On the other hand, PULP Trainlib~\cite{trainlib}, Cioflan \etal~\cite{cristi}, and Ravaglia \etal~\cite{ravaglia} are all examples of approaches to enable on-device learning and adaptation on \riscv multi-core \gls{pulp} clusters like Vega~\cite{vega}, that provide mixed \gls{fp} precision capabilities, spanning from IEEE 754 Standard \gls{fp}32 and \gls{fp}16 to \textit{bfloat}. However, the low speed and number of available floating point units typical of ultra-low-power microcontrollers limit the performance of these libraries.

\subsection{Training Accelerators}
To address the limited training performance achievable by software libraries running on low-power processors, several researchers turned to hardware acceleration~\cite{9926331}. 

Cambricon-Q~\cite{cambricon} is a training-oriented chip for high accuracy and energy efficiency based on \dbit{8} fixed-point arithmetic. However, many common training algorithms require floating-point operations to ensure convergence~\cite{8bit_trainig, 9515082,fox2021a, NEURIPS2018_335d3d1c}. Most training-oriented chips employing \gls{fp} arithmetic are all characterized by power envelopes unsuitable for extreme-edge applications. IBM proposes~\cite{4c_ibm, rapid}, an AI computing platform featuring $8\times8$ mixed-precision engines supporting \gls{fp}16 and hybrid \gls{fp}8 training, while~\cite{oh_ibm} support only \gls{fp}16 and \gls{fp}32. Similarly, LNPU~\cite{lnpu} supports mixed \dbit{8} and \dbit{16} \gls{fp} precision for on-chip training. While these chips consume significantly less power than data-center \glspl{gpu} during \gls{nn} training (i.e. a few Watts as opposed to hundreds of Watts), they still do not meet the tens of mW power constraints of TinyML devices.

Recently, a few training-oriented \glspl{soc} that fit the power budget of extreme-edge applications have been presented. T-PIM~\cite{tpim} is a processing-in-memory accelerator in \SI{28}{nm} technology for on-device learning. It reaches up to \SI{250}{GOPS/W} during training with $0\%$ of sparsity and within a power envelope of \SI{51.23}{mW} at \SI{280}{MHz} operating frequency. However, T-PIM and all the recently proposed PIM approaches do not support \gls{fp} computations and are not suitable for standard back-propagation. To support \glspl{nn} training at reduced power budgets, many training-oriented chips extensively employ pruning to increase sparsity during training~\cite{albus}, lacking generality. For example, TSUNAMI~\cite{tsunami} and Trainer~\cite{trainer} are both accelerators designed for extreme-edge \gls{nn} inference and training, meeting the TinyML power constraints by employing pruning and zero skipping.  Anders~\etal~\cite{anders} propose a reconfigurable accelerator for dense-sparse matrix multiplications for mixed-precision computations, suitable for training-oriented applications since it features \gls{fp}16 multiplications and \gls{fp}32 accumulations with low area occupation and high energy efficiency. However, such an accelerator is not parametric, thus not allowing a fast scale-up at design time when higher performance is needed. In addition, its integration into a real system has not been evaluated, and it does not support compressed \gls{fp}8 input/output tensors, which allows for training larger NN models on edge devices where the memory resources are limited.

\begin{table}[t]
\centering
\footnotesize
\caption{Set of General Matrix-Matrix Operations supported by RedMulE}
\label{tab:gemmops-table}
\begin{tabular}{|ccccc|}
\hline
\multicolumn{5}{|c|}{\textbf{$\mathbf{Z = (X \circ W) \star Y}$}} \\ \hline
\multicolumn{1}{|c|}{\textbf{Group}} &
  \multicolumn{1}{c|}{\textbf{Kernel}} &
  \multicolumn{1}{c|}{\textbf{$\mathbf{\circ}$}} &
  \multicolumn{1}{c|}{\textbf{$\mathbf{\star}$}} &
  \textbf{Res} \\ \hline
\multicolumn{1}{|l|}{} &
  \multicolumn{1}{c|}{Matmul} &
  \multicolumn{1}{c|}{$\times$} &
  \multicolumn{1}{c|}{$+$} &
  $Z = (X \times W) + Y$ \\ \hline
\multicolumn{1}{|c|}{\multirow{4}{*}{\textbf{\begin{turn}{90}Group 1\end{turn}}}} &
  \multicolumn{1}{c|}{\begin{tabular}[c]{@{}c@{}}Maximum\\ Critical Path\end{tabular}} &
  \multicolumn{1}{c|}{$+$} &
  \multicolumn{1}{c|}{$max$} &
  $Z = max[Y, (X + W)]$ \\ \cline{2-5} 
\multicolumn{1}{|c|}{} &
  \multicolumn{1}{c|}{\begin{tabular}[c]{@{}c@{}}All-Pairs\\ Shortest Paths\end{tabular}} &
  \multicolumn{1}{c|}{$+$} &
  \multicolumn{1}{c|}{$min$} &
  $Z = min[Y, (X + W)]$ \\ \cline{2-5} 
\multicolumn{1}{|c|}{} &
  \multicolumn{1}{c|}{\begin{tabular}[c]{@{}c@{}}Maximum\\ Reliability Path\end{tabular}} &
  \multicolumn{1}{c|}{$\times$} &
  \multicolumn{1}{c|}{$max$} &
  $Z = max[Y, (X \times W)]$ \\ \cline{2-5} 
\multicolumn{1}{|c|}{} &
  \multicolumn{1}{c|}{\begin{tabular}[c]{@{}c@{}}Minimum\\ Reliability Path\end{tabular}} &
  \multicolumn{1}{c|}{$\times$} &
  \multicolumn{1}{c|}{$min$} &
  $Z = min[Y, (X \times W)]$ \\ \hline
\multicolumn{1}{|c|}{\multirow{2}{*}{\textbf{\begin{turn}{90}Group 2\end{turn}}}} &
  \multicolumn{1}{c|}{\begin{tabular}[c]{@{}c@{}}Minimum\\ Spanning Tree\end{tabular}} &
  \multicolumn{1}{c|}{$max$} &
  \multicolumn{1}{c|}{$min$} &
  $Z = min[Y, max(X, W)]$ \\ \cline{2-5} 
\multicolumn{1}{|c|}{} &
  \multicolumn{1}{c|}{\begin{tabular}[c]{@{}c@{}}Maximym\\ Capacity Path\end{tabular}} &
  \multicolumn{1}{c|}{$min$} &
  \multicolumn{1}{c|}{$max$} &
  $Z = max[Y, min(X, W)]$ \\ \hline
\end{tabular}
\end{table}

\subsection{GEMM-Ops Chips}
All examples of training and inference-oriented chips mentioned so far target only the most common \gls{dl} operations (such as matrix multiplications and convolutions). However, a large set of kernels share the same computational structure as \gls{gemm} but do not rely on multiplication and addition as elementary operations, falling into the \glspl{gemmop} scope. Graph analytics, such as breadth-first search~\cite{j_gilbert1, j_gilbert2}, short-distance problems~\cite{semiring} that are commonly used for path planning optimization in embedded drones navigation~\cite{kamp}, and minimum spanning tree, used for computer vision~\cite{minspantree}, are examples of applications that make use of \glspl{gemmop}. SIMD\textsuperscript{2}~\cite{simd2} addresses this issue by building functional units for \gls{gemmop} acceleration on top of NVIDIA Streaming Multiprocessor architecture, resembling the TensorCores structure and providing dedicated ISA extensions. The design is implemented in \SI{45}{nm} technology. Adding all the SIMD\textsuperscript{2} extensions to the baseline matrix multiplication unit results in up to $15.8\times$ speedup with respect to executing the same kernel on CUDA cores at the cost of $69\%$ of area overhead.

In this paper, we propose an extended version of RedMulE~\cite{redmule} with the following unique combination of features:
\begin{itemize}
    \item An array of Floating-Point Units-based Computing Elements (CEs) for efficient training and inference of general \gls{dl} models on embedded \glspl{soc} with additional support for reduced bit-width \gls{fp} computation. We tightly couple RedMulE with a parallel cluster of RISC-V processors to achieve maximum flexibility in implementing complex training algorithms;
    \item Supports for \glspl{gemmop} with a low area overhead ($16\%$) with respect to a \gls{gemm}-only implementation to address a wider spectrum of applications;
    \item A fully-parametric design that allows the instantiation of a wide range of CEs arrays, internal buffers and memory interface configurations.
\end{itemize}

\section{Background} \label{section:background}

\subsection{Generalized Matrix-Matrix Operations}

In this work, we define \textit{Generalized Matrix-Matrix Operations (GEMM-Ops)} as all the operations of the kind $f2(\mathbf{Y}, f1(\mathbf{X}, \mathbf{W}))$, in particular they can be expressed as:

\vspace{-2mm}
\begin{equation}\label{eq:1}
    \mathbf{Z}=(\mathbf{X} \circ \mathbf{W}) \star \mathbf{Y}
\end{equation}

where $\circ$ corresponds to $f1()$ and $\star$ corresponds to $f2()$. Table~\ref{tab:gemmops-table} shows some examples of \glspl{gemmop}, divided into two groups. Group 1 includes all the \glspl{gemmop} where the $\circ$ operator can be of the $+/\times$ kind while $\star$ can be $min/max$. Group 2 contains the \glspl{gemmop} kernels where the $\circ$ operator also belongs to the $min/max$ kind. $\mathbf{X}$ is a matrix of size $M\times N$, $\mathbf{W}$ is a matrix of size $N\times K$, while $\mathbf{Z}$ and $\mathbf{Y}$ have size $M\times K$.

\begin{figure}[t]
    \centering
    \includegraphics[width=0.9\linewidth]{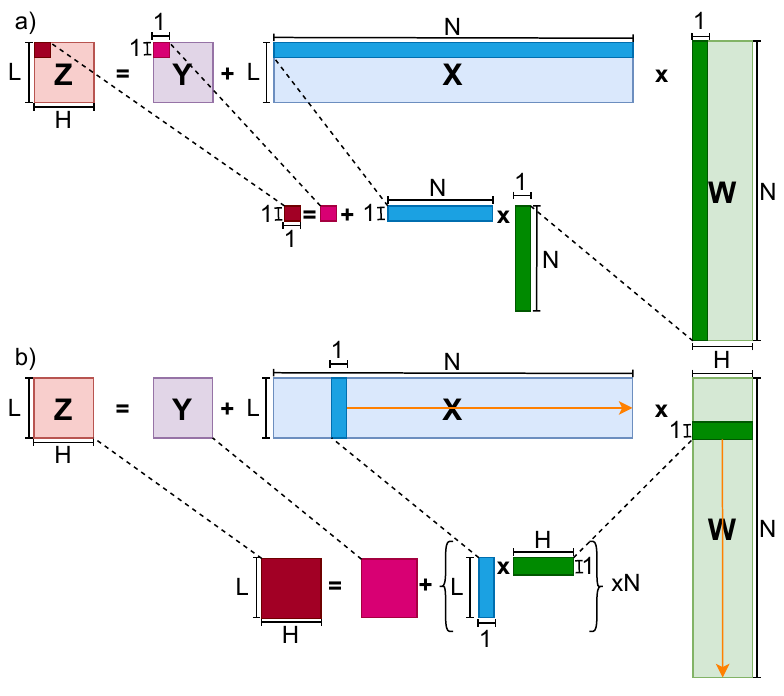}
    \vspace{-2mm}
    \caption{Execution of a \gls{gemm} through a) scalar dot product microkernel and b) block-dot product (or outer product) microkernel.}
    \label{fig:blas}
    \vspace{-4mm}
\end{figure}

The similarity of \glspl{gemm} and \glspl{gemmop} makes matrix computing units good candidates to be extended for supporting \glspl{gemmop}, extending their flexibility to accelerate generalized parallel algebraic operators. This class of algorithms is also well-suited for \gls{ml} applications since matrices are the baseline structure of all \gls{dl} models. To this purpose, it is essential to note that the structure of Equation~\ref{eq:1} is symmetric. As a consequence, for \gls{ml} applications, there is no need to identify $\mathbf{X}$ or $\mathbf{W}$ as input or weight matrices because their role can be flexibly exchanged.

\subsection{Asymptotic Optimality of Linear Algebra Acceleration Strategies} \label{subsection:efficient_hw_design}
Memory load/store operations enlarge the gap between theoretical and practical performance and efficiency. Therefore, maximizing the number of operations performed per memory access, i. e. the arithmetic intensity, is the key to an efficient design. As analyzed by Pedram~\cite{pedram}, scalar dot products and vector units do not guarantee the best trade-off between the number of operations performed per memory load/store access. As shown in Fig.~\ref{fig:blas}a, a simple scalar dot product that operates on a $N$-dimensional array performs $2 \times N$ operations ($N$ multiplications + $N$ additions). The memory operations performed in this kernel are $N$ loads of $X$, $N$ loads of $W$, one load of $Y$ and one store of $Z$. The resultant arithmetic intensity is:

\vspace{-2mm}
\begin{equation}\label{eq:2}
    Intensity\_1D = \frac{OPs}{LD/ST} = \frac{2N}{2N + 2} \sim 1,\quad (N\rightarrow\infty).
\end{equation}

2-Dimensional $L \times H$ arrays exploit block-dot products (outer product) microkernels to perform \glspl{gemm}. Let us consider an $L \times H$ 2D array that can operate on $L \times 1$ and $1 \times H$ vectors, each made of $N$ elements, like those shown in Fig.~\ref{fig:blas}b. The operations performed on the two vectors are $2 \times L \times H$, repeated $N$ times. The resulting load/store operations are $L \times N$ loads of $X$, $H \times N$ loads of $W$, $L \times H$ loads of $Y$ and $L \times H$ stores of $Z$. With these changes, Equation~\ref{eq:2} becomes:

\vspace{-2mm}
\begin{equation}\label{eq:3}
    \frac{OPs}{LD/ST} = \frac{2LHN}{(L+H)N + 2LH} \sim \frac{2LH}{L+H},\quad (N\rightarrow\infty).
\end{equation}

Equation~\ref{eq:3} shows that if $L=H$, the number of operations is quadratic with the size of the 2-D array, while the number of memory accesses remains linear. This demonstrates that 2-dimensional arrays are more efficient with respect to scalar or vector units. Thus, we will exploit the outer-product approach for the RedMulE design.

\section{Architecture}\label{section:architecture}

\begin{figure}[t]
    \centering
    \includegraphics[width=0.7\linewidth]{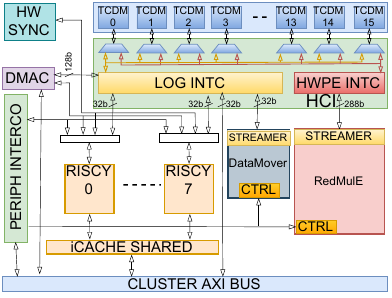}
    \vspace{-2mm}
    \caption{PULP cluster architecture with HWPEs integration.}
    \label{fig:pulp_archi}
    \vspace{-4mm}
\end{figure}

In this section, we describe the \gls{pulp} cluster, the hardware template we rely upon, and the RedMulE micro-architecture.

\subsection{PULP Cluster and RedMulE}
In Fig.~\ref{fig:pulp_archi}, we show the architecture of a \gls{pulp} cluster, a multi-core architecture that features a parametric number (2--16) of \dbit{32} \riscv general-purpose cores featuring a partially shared, partially private instruction cache. In this specific work, we focus on a \gls{pulp} cluster containing 8 \riscv cores, equipped with \SI{128}{kB} of \gls{tcdm} split among 16 banks for word-level interleaving with a low level of contention. The \gls{pulp} cluster also features an event unit for flexible internal synchronization and a dedicated \gls{dmac} to efficiently move data between the \gls{tcdm} and external memories. A peripheral interconnect allows the \riscv cores to program the on-board peripherals (like the \gls{dmac}), and an AXI4 full cross-bar interconnect allows communications with the external environment. The \gls{pulp} cluster also features a system-level clock gating cell to eliminate its dynamic power consumption when the cluster is not in use.

The capabilities of the \gls{pulp} cluster can be further enhanced by integrating application-specific hardware accelerators called \glspl{hwpe}. \glspl{hwpe} are software programmed by the \riscv cores through the peripheral interconnect and share the \gls{tcdm} with the \riscv cores and the \gls{dmac}. In this sense, the \glspl{hwpe} are tightly-coupled with the cluster cores~\cite{xnor}. The cores, the \gls{dmac}, and the accelerators access the shared \gls{tcdm} through a single-cycle latency \gls{hci}~\cite{darkside_j}. Such interconnect features a \textit{logarithmic} branch that allows all-to-all single-cycle accesses from \dbit{32} master ports, like those of the cores or the \gls{dmac}, to each of the word-interleaved memory banks. Conflicts are managed by granting only one initiator per bank with a round-robin scheme. The other branch is the \textit{shallow} branch. It features a single \dbit{n} parametric port, routed to adjacent \dbit{32} memory banks treated like a single wider bank without arbitration. This branch allows for simple integration of tightly-coupled accelerators like \glspl{hwpe}. The bitwidth of the shallow branch port can be tuned to the \gls{hwpe} requirements through a parameter. The \gls{tcdm} banks are connected to the two \gls{hci} branches through a set of multiplexers, which grant access to one branch or the other according to a configurable starvation-free rotation scheme, allocating a configurable maximum of K $<$ N consecutive cycles to the \gls{hwpe} over a period of N cycles.

\begin{figure*}[t]
    \centering
    \includegraphics[width=\textwidth]{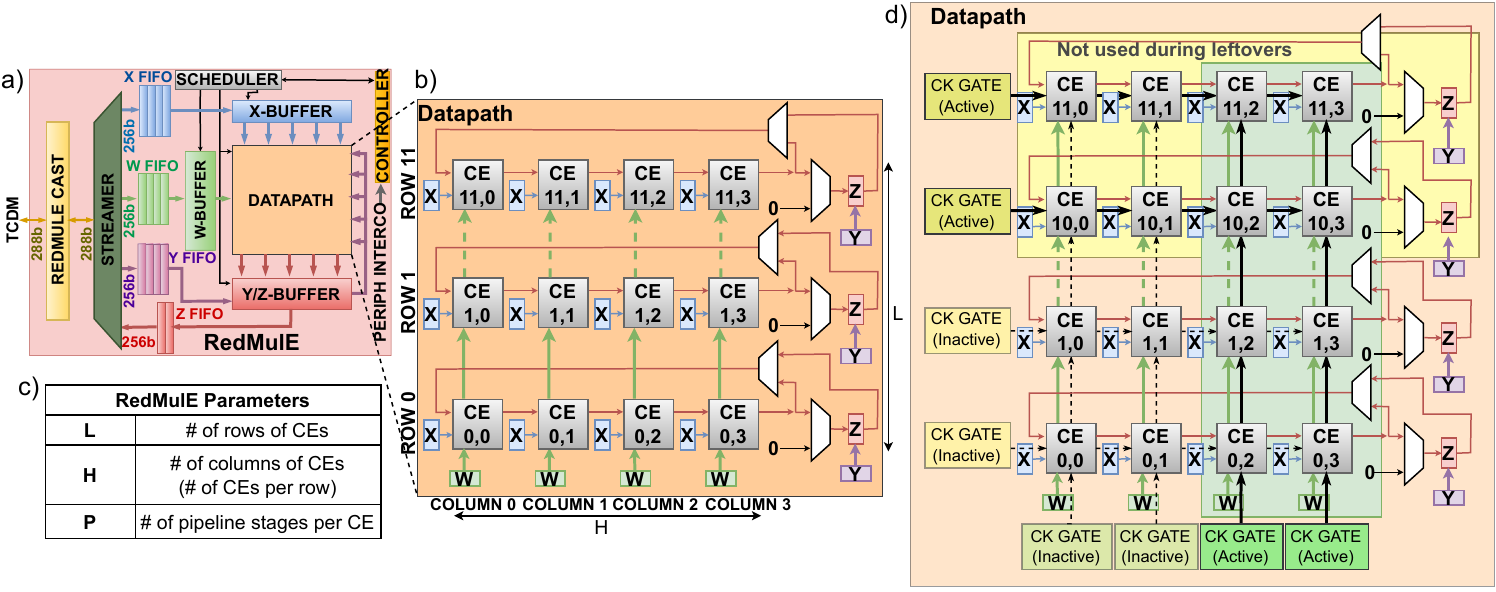}
    \vspace{-2mm}
    \caption{a) RedMulE internal architecture; b) RedMulE \textit{Datapath} microarchitecture; c) Table with RedMulE design-time available parameters; d) Datapath-level clock gating logic.}
    \label{fig:datapath}
    \vspace{-4mm}
\end{figure*}

During the execution of \glspl{nn} workloads, particularly during inference and training, on-the-fly data marshalling operations are known to reduce performance significantly. For this reason, our \gls{pulp} cluster features a DataMover~\cite{darkside_j}. The DataMover is a tiny accelerator capable of transposing 3-dimensional tensors stored in the \gls{tcdm}, with $33\%$ less time than eight \riscv cores and up to 50$\times$ increased energy efficiency (the lower the precision of chunks to transpose the more significant the advantages). The accelerator works with a configurable data element bitwidth, from \dbit{32} down to \dbit{1}.

\subsection{RedMulE}
\subsubsection{Global Architecture}
RedMulE is a domain-specific processor designed to accelerate \glspl{gemmop}. Its architecture is shown in Fig.~\ref{fig:datapath}a. The core of RedMulE is the \textit{Datapath}, a 2-Dimensional array of \glspl{ce} interconnected as shown in Fig.~\ref{fig:datapath}b. The \glspl{ce} are organized in \textit{L} rows, each made of \textit{H} columns. Within each row, a number of \textit{H} \glspl{ce} are cascaded so that each \gls{ce} computing an intermediate product will pass its result to the next \gls{ce}. The partial product computed by each row's last \gls{ce} is fed back as accumulation input of the same row's first \gls{ce}. The RedMulE \textit{Datapath} features a design-time configurable number of internal \glspl{ce}, pipeline registers (\textit{P}) for each \gls{ce}, and internal computing precision (\gls{fp} bitwidth). All RedMulE's parameters are tunable at design time and are resumed in Fig.~\ref{fig:datapath}c.

\begin{figure}[t]
    \centerline{\includegraphics[width=0.8\linewidth]{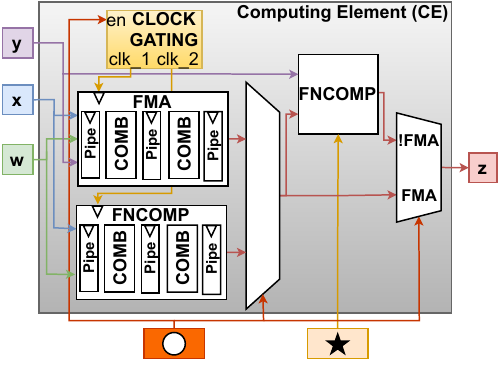}}
    \caption{Microarchitecture of RedMulE's \gls{ce} with extensions for \glspl{gemmop} support}
    \label{fig:comp-elem}
\end{figure}

Fig.~\ref{fig:datapath}d shows part of the hierarchical clock gating logic implemented within RedMulE. The proposed power-saving scheme leverages a fine-grained clock-gating mechanism that reduces the active power consumption of unused portions (rows and columns) of RedMulE’s \textit{Datapath}. RedMulE can dynamically disable the rows of \glspl{ce} that would remain unused during the calculations involving the leftovers. The same holds for the columns of \glspl{ce}, where RedMulE gates the clock of a given column within the \textit{Datapath} depending on the calculation phase. 
Within the \textit{Datapath}, each \gls{ce} is divided into two stages, each dedicated to selecting the $\circ$ and $\star$ operations shown in Table~\ref{tab:gemmops-table}. Each \gls{ce} relies on computing \gls{fp} modules adapted from the open-source \textit{FPnew} trans-precision \gls{fpu}~\cite{fpnew} so that their internal pipeline registers could support backpressure coming from memory stalls during RedMulE's operation. Fig.~\ref{fig:comp-elem} and Section~\ref{section:ce-uarc} will describe in detail how the proposed clock gating scheme hierarchically propagates within the \gls{ce}'s micro-architecture, where dedicated clock gating cells are in charge of disabling the unused units.

To feed the \textit{Datapath} with data, RedMulE includes the \textit{Streamer}, following the \gls{hwpe} design strategy~\footnote{https://hwpe-doc.rtfd.io}. The \textit{Streamer} is a specialized memory access unit that connects RedMulE to the \gls{hci} shallow branch through a single wide port of parametric size (multiple of \dbit{32}), used for load and store operations. The incoming stream from the \gls{hci} is propagated to a single input-multiple output dispatcher that forwards the valid only to the selected output channel; simultaneously, each output channel propagates the incoming stream from the \gls{hci} to the accelerator input ports. On the other hand, the streams produced by RedMulE are propagated to the \gls{hci} interface during write operations.

The \textit{Streamer} is connected to three internal buffers: an X-Buffer that changes all the \textit{L} inputs of a column once every $H\times(P+1)$ cycles; a W-Buffer made of \textit{H} shift registers, each broadcasting a new $\mathbf{W}$-element to all the \textit{L} \glspl{ce} of a column every cycle; a Z-Buffer that buffers the computed $\mathbf{Z}$-elements. The same buffer is used to pre-load $\mathbf{Y}$-elements and push them into the \textit{Datapath}. This solution saves area and power in the accelerator since there is no need for a separated buffer to store $\mathbf{Y}$ bias. All three buffers benefit from the clock gating logic shown in Fig.~\ref{fig:datapath}d. In particular, during the calculations of leftovers in input and output matrices, some of the rows of the \textit{Datapath} do not perform any computation, and the corresponding lines in the \textit{X} and \textit{Z} buffers are clock gated as well. The same consideration holds for the columns of \glspl{ce} and the lines in the \textit{W} buffer. If a column of \glspl{ce} does not perform any computation due to leftovers' presence, the corresponding line in the \textit{W}-buffer is dynamically clock gated.

The control side of the accelerator is divided into two sub-modules, namely \textit{Scheduler} and the \textit{Controller}, that contain the register file, accessed by the cores to program the accelerator and cooperate to regulate the accelerator execution.

\subsubsection{GEMM-Ops Extension to the Computing Element}\label{section:ce-uarc}
We extended each \gls{ce} of RedMulE with dedicated hardware to support the execution of all the \glspl{gemmop} in Table~\ref{tab:gemmops-table}. Fig.~\ref{fig:comp-elem} shows the microarchitecture of each \gls{ce}, which features two stages. The first stage selects the $\circ$ operation and contains one \gls{fma} unit and one \gls{fncomp} unit which implements \gls{fp} MIN/MAX operations. $\mathbf{X}$ and $\mathbf{W}$ elements are then propagated to both the \gls{fma} and \gls{fncomp} modules. Depending on the desired $\circ$ operation, a multiplexer selects the result of either the \gls{fma} or the \gls{fncomp}. At the same time, the clock gating module shown in Fig.~\ref{fig:comp-elem} freezes the input operands of the unused modules to completely eliminate undesired switching activity and related dynamic power. The \gls{fma} and \gls{fncomp} feature \textit{P} pipeline registers each, guaranteeing that the two modules introduce the same latency. We adapted the pipeline of both \gls{fma} and \gls{fncomp} modules from the open-source \textit{FPNew} trans-precision \gls{fpu}~\cite{fpnew} to make them capable of supporting back pressure coming from memory stalls during RedMulE's operation.

In each \gls{ce}, the $\mathbf{Y}$ element is propagated to the \gls{fma} unit in the first stage and is also directly connected to the input of the second stage of the \gls{ce}. The second stage of each \gls{ce} allows the $\star$ operation selection and contains a fully combinational \gls{fncomp} module. The output multiplexer allows choosing the output of the second stage \gls{fncomp}, in case of \gls{gemmop}, or that of the first stage \gls{fma} if a simple \gls{gemm} operation is performed. The proposed architectural solution guarantees the execution of all the operations listed in Table~\ref{tab:gemmops-table} with a compact implementation, in which we add just what is strictly needed.

\subsubsection{Mixed-Precision Extension}
Hybrid \gls{fp}8 precision formats can be used as an efficient compression scheme to enable \gls{dl} inference and training on extreme-edge devices. Hybrid \gls{fp}8 precision means that the $\{sign, exponent, mantissa\}$ structure used to represent the tensors can be either $\{1,5,2\}$ or $\{1,4,3\}$\footnote{Also called E4M3 and E5M2 by NVIDIA [https://tinyurl.com/mkhbxj3v]}. The former format is best suited for backward propagation of gradients, as it provides a larger dynamic range but a lower accuracy. At the same time, the latter is a better fit for forward propagation of activations thanks to the larger mantissa~\cite{hyb_fp8, fp8_train}. While \dbit{8} representation effectively reduces data footprint, it could severely impact accuracy due to reduced precision accumulations. To support this use case, RedMulE works internally with fixed \gls{fp}16 precision but still accepts compressed \gls{fp}8 formats as inputs and can generate \gls{fp}8 compressed output tensors. To do this, we extended RedMulE’s architecture with a dedicated casting module placed between the Streamer and the \gls{hci} interface, as shown in Fig.~\ref{fig:datapath}a.

\begin{figure}[t]
    \centerline{\includegraphics[width=0.7\linewidth]{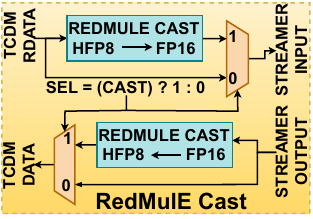}}
    \caption{Internal architecture of RedMulE's cast module for mixed precision support.}
    \label{fig:cast-module}
\end{figure}

Fig.~\ref{fig:cast-module} shows the architectural implementation of the casting module. It contains two \gls{fp} cast units: the input one casts \dbit{8} \gls{fp} incoming streams into \dbit{16} \gls{fp} to feed the accelerator so that the \glspl{ce} in the Datapath can operate on larger precision, guaranteeing high accuracy during intermediate accumulations. After the computation, the output cast unit converts the \dbit{16} \gls{fp} results produced by RedMulE to \dbit{16} or \dbit{8} encoded outgoing streams before writing it to memory. The cast units can be excluded from the path if the input tensors are represented with 16 bits.

For \gls{dl} use cases only, RedMulE can also be instantiated at design time to only load and store H\gls{fp}8 operands. In this use-case, the input and output tensors represented with \dbit{8} formats allow to read or write from and to the memory twice the number of elements while keeping the same memory bandwidth. Consequently, this allows for doubling the number of \glspl{ce} inside each row, doubling RedMulE's performance compared to the \dbit{16} inputs case.

\begin{figure*}[t]
    \centerline{\includegraphics[width=\textwidth]{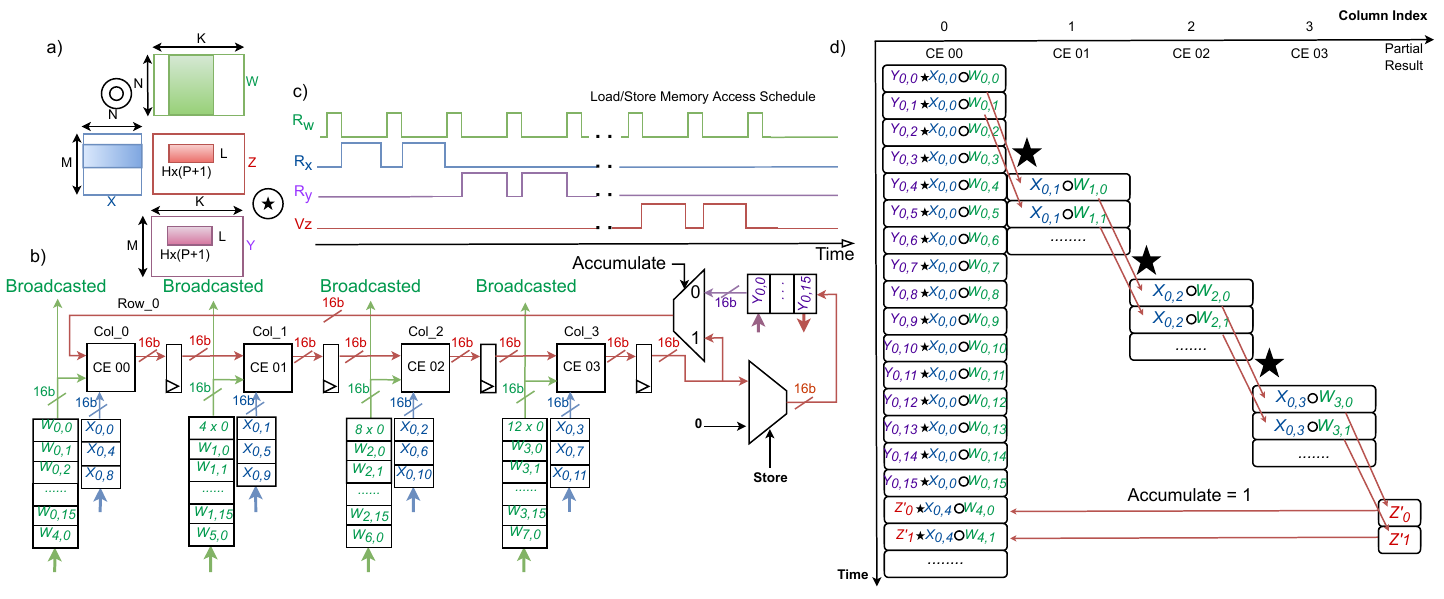}}
    \caption{a) \gls{gemmop} execution displayed on matrices; b) Row of \glspl{ce} within RedMulE \textit{Datapath}; c) Memory access schedule in load/store mode described in terms of \textit{R} (Ready) and \textit{V} (Valid) handshake signals; d) Pipeline evolution within a row of ces.}
    \label{fig:schedule}
\end{figure*}

\subsection{RedMulE Computational Model}
Fig.~\ref{fig:schedule}a shows how RedMulE performs a \gls{gemmop} visualising it on the computed matrices, while Fig.~\ref{fig:schedule}b and Fig.~\ref{fig:schedule}d show the detailed sequence of the operations within a row of \glspl{ce} providing an example of \gls{gemm} execution. For this discussion, let us focus on a RedMulE implementation that features $L=12$, $H=4$, and $P=3$.
The RedMulE operation starts by pre-loading the Z-Buffer with \textit{L} rows from the $\mathbf{Y}$-matrix, each row made of $H \times (P+1) = 16$ \gls{fp}16 elements (\dbit{256} memory width/\dbit{16} internal precision), namely $\mathbf{y _{0,0}}$ - $\mathbf{y_{0,15}}$ for Row\_0, $\mathbf{y_{1,0}}$ - $\mathbf{y_{1,15}}$ for Row\_1, and so on. Afterwards, RedMulE pre-loads the X-Buffer as well, following the same pattern, and then loads a set of $H \times (P+1) = 16$ $\mathbf{W}$-elements ($\mathbf{w_{0,0}}$ - $\mathbf{w_{0,15}}$) inside the first shift register of the W-buffer. Each $\mathbf{W}$-element is broadcasted to all the \textit{L} \glspl{ce} in the first \textit{Datpath} column. While $\mathbf{W}$-elements are broadcasted, the Z-Buffer pushes $\mathbf{Y}$-elements in the \glspl{ce} array cycle-by-cycle to perform the $\star$ operation during the execution of the $\circ$ one.

After $P+1$ cycles, each of the \textit{L} \glspl{ce} in the first column forwards its computed partial result to the neighbour \gls{ce} in the second column. The accelerator loads another set of $H \times (P+1)$ $\mathbf{W}$-elements ($\mathbf{w_{1,0}}$ - $\mathbf{w_{1,15}}$) to broadcast them to all the \glspl{ce} in the second column. Once all the \textit{H} \glspl{ce} of a row have completed their computations, calculating a subset of $H \times (P+1)$ row-column intermediate results, RedMulE activates its feedback ($accumulate = 1$) to provide the intermediate results to the accumulation input of the first \glspl{ce} of the given row, then reiterating the computation. Immediately after, the \textit{Streamer} reloads the next $\mathbf{Y}$-submatrix in the Z-Buffer so that it will be ready for the next calculation. During the Z-Buffer reload operation, the X-Buffer provides a new $\mathbf{X}$-operand to the first column of \glspl{ce}, and a new set of $H \times (P+1)$ $\mathbf{W}$-elements is reloaded in the first W shift register. After $(P+1)$ cycles, all the \textit{L} \glspl{ce} of the first column produce a new partial product and provide it to the \glspl{ce} in the second column. The X-Buffer provides a new $\mathbf{X}$-operand at the input of the second column of \glspl{ce}, and the W-Buffer loads a new set of $H \times (P+1)$ $\mathbf{W}$-elements in the second W shift register for broadcasting, and the computation continues. Fig~\ref{fig:schedule}d shows the detailed sequence of data within the pipeline of a row of \glspl{ce} from the beginning of a \gls{gemm} operation until the moment of the reuse of the partial results ($accumulate = 1$).

\vspace{-1mm}
To guarantee a continuous data flow in the accelerator, the W-buffer accesses the memory once every $(P+1)$-cycles to load a new set of $H \times (P+1)$ $\mathbf{W}$-elements. Once the X-Buffer and the Z-Buffer are empty, RedMulE reuses the \textit{Streamer} port to load the $\mathbf{X}$ and $\mathbf{Y}$-operands. Such operation is made by interleaving the memory accesses to $\mathbf{X}$ or $\mathbf{Y}$ matrices between two adjacent $\mathbf{W}$-matrix accesses until the complete fulfilment of the X and Z buffers. Fig~\ref{fig:schedule}c shows how the memory accesses to different matrices are interleaved, describing the memory accesses in terms of Ready (\textit{R}) and Valid (\textit{V}) handshake signals. The \textit{Streamer} load and store units fully support backpressure through a mechanism based on \textit{R}/\textit{V} handshake signals. Such a mechanism fully decouples the memory access and data consumption/production from the \textit{Datapath}. The \textit{V} signals for loads and the \textit{R} signals for stores are generated within the \textit{Streamer} itself depending only on memory stalls, which can be amortized by the presence of FIFO elements, and not on the actual usage from the \textit{Datapath}. On the other hand, the \textit{Datapath} uses the \textit{R} signal of loads and the \textit{V} signal of stores, as shown in Fig.~\ref{fig:schedule}c, to control the order of memory accesses interleaving them so that a continuous dataflow can be maintained. This choice is made to maximize the memory port utilization since having a single memory port also helps reduce the overall streamer area.

After the conclusion of an entire row-column operation, the Z-Buffer buffers the final sub-matrices. Afterwards, store operations are interleaved between two adjacent $\mathbf{W}$ load accesses until the Z-Buffer is empty and can be reloaded with $\mathbf{Y}$-elements. With this approach, RedMulE optimizes the bandwidth utilization using a single wide memory port and achieves up to $99.4\%$ \glspl{ce} utilization.

\section{Implementation and Measurements} \label{section:measurement}

\begin{figure*}[t]
    \centering
    \includegraphics[width=\textwidth]{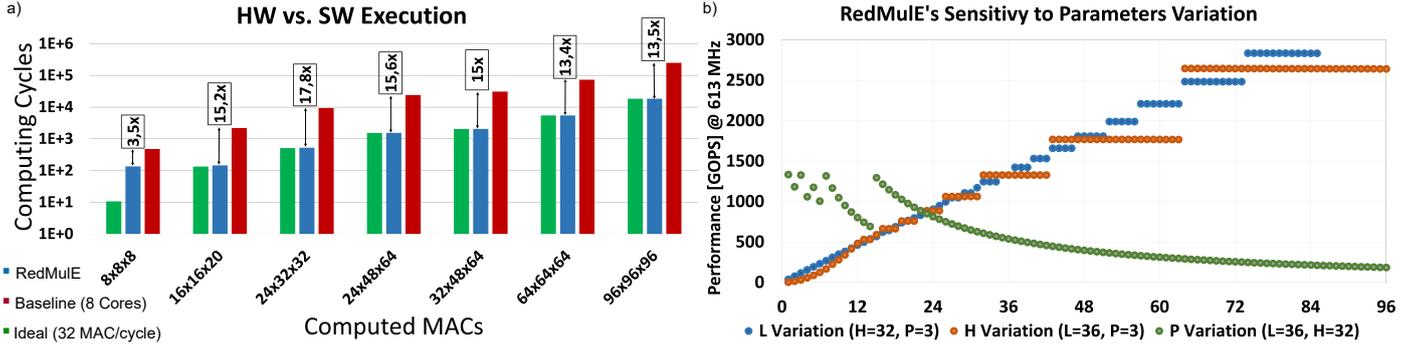}
    \vspace{-2mm}
    \caption{a) RedMulE's synthetic \gls{gemm} execution compared with software executed on 8 \riscv cores; b) RedMulE's sensitivity to L, H, and P parameters.}
    \label{fig:gemm-eval}
\end{figure*}

\subsection{Experimental Setup}
We focus our experiments on a RedMulE\textsubscript{12x4} instance with $H=4$, $L=12$, $P=3$, resulting in 48 \glspl{ce} and a \dbit{288} wide \gls{hci} port, for \dbit{256} + \dbit{32} non-word-aligned accesses. We also address a RedMulE\textsubscript{12x8} since, as described in Section~\ref{section:HFP8_train}, it uses the same memory interface with twice the number of \glspl{ce}.

Our experiments target GlobalFoundries \SI{22}{nm} technology using Synopsys Design Compiler for synthesis (slow corner at $f_\mathrm{targ}=\SI{250}{MHz}$, $V_{DD}=\SI{0.59}{V}$, $T=\SI{125}{\celsius}$) and Cadence Innovus for full-cluster Place\&Route in the same operating point. RedMulE's timing analysis and power extraction were made using Prime Time with $100\%$ annotated switching activity from post-layout simulation in typical corner at $\SI{25}{\celsius}$, targeting two operating points: \SI{470}{MHz} at \SI{0.65}{V} for high energy efficiency and \SI{613}{MHz} at \SI{0.8}{V} for high performance.

\subsection{Performance Evaluation}
\subsubsection{GEMM Performance Evaluation}
We use square and rectangular matrices as a synthetic benchmark to evaluate RedMulE's computation latency in cycles against the SW execution on 8 parallel \riscv cores sharing 4 \glspl{fpu}. On the given benchmark, RedMulE reaches a peak throughput of more than \SI{95.4}{OP/cycle}, where we count both $\star$ and $\circ$ as one "OP", e.g. for a regular \gls{gemm} we count 1 MAC = 2 OPs. RedMulE achieves up to $99.4$\% of \glspl{ce} utilization on $96\times96$ \gls{fp}16 matrices (55 kB memory occupation), leading to \SI{58.5}{GFLOPS} at \SI{613}{MHz} with \SI{0.80}{V} supply. Fig.~\ref{fig:gemm-eval}a shows the number of computing cycles required to compute various matrices during parallel \gls{fp}16 software executed on 8 \riscv cores and compares them on RedMulE, showing that it reaches $15\times$ average speedup over the software on large matrices. This performance increase with respect to the software counterpart settles around $13\times$ with larger matrices since also the software execution becomes more efficient in those cases. We also consider the acceleration of a small $8 \times 8 \times 8$ case, as shown in Fig.~\ref{fig:gemm-eval}a in which the accelerator is under-utilized, but it still introduces $3.5\times$ speedup over the software parallel execution.

We evaluated the sensitivity of RedMulE to the \textit{L}, \textit{H}, and \textit{P} parameters variation by executing a GEMM kernel with dimensions fixed to \textit{M}=512, \textit{N}=512, and \textit{K}=512. We show the effect of parameters' variation in Fig.~\ref{fig:gemm-eval}b. Each RedMulE tile features \textit{L} rows and $H\times(P+1)$ columns. For this reason, each computation of RedMulE accesses the $\mathbf{X}-matrix$ a number of times equal to $\frac{M}{L}\times\frac{N}{H\times(P+1)}$, while it accesses the $\mathbf{W}$-matrix a number of times equal to $N\times\frac{K}{H\times(P+1)}$. Looking at the accelerator behaviour when sweeping the \textit{L} parameter (the number of rows within the RedMulE datapath), considering that each tile of the matrix depends on the ratio $\frac{M}{L}$ (number or rows of the $\mathbf{X}$-matrix divided by the number of rows in the accelerator), the performance of RedMulE rises linearly until the tile ratio starts to flatten. When this happens, the performance behaviour starts stepping up every time an increase of the \textit{L} value allows to reduce the number of iterations on the $\mathbf{X}$-matrix. Then it flattens again until the next step happens.

Similar considerations hold for the sweep of the \textit{H} parameter. Since RedMulE's accesses to both $\mathbf{X}$-matrix and $\mathbf{W}$-matrix depend on $H\times(P+1)$, as \textit{H} increases, the performance of RedMulE increases until the ratio $\frac{N}{H\times(P+1)}$ and $\frac{K}{H\times(P+1)}$ flattens. Then, the performance curve steps up every time an increase of the \textit{H} parameter reduces the number of accesses to the $\mathbf{X}$-matrix and the $\mathbf{W}$-matrix before flattening again. Since the value of \textit{L} is upper-bounded by $H\times P$, the value of \textit{L} also changes in the first part of the sweep accordingly with the $H\times P$, explaining the initial non-linear behaviour of the curve.

Compared with the previous cases, the performance trend changes when sweeping the \textit{P}. Increasing the number of pipeline registers not only traduces into larger internal tiles and thus reduces the number of times RedMulE accesses the $\mathbf{X}$ and $\mathbf{W}$ matrices, but also turns into a higher internal capacity of the accelerator. By keeping the input matrices' dimensions fixed, RedMulE's performance drops as pipeline registers increase because progressively fewer data are available to feed the internal pipeline queues. Consequently, the leftovers and the computation time increase, while the performance drops due to higher latency introduced by the pipeline registers. When the value of \textit{P} creates a tile that reduces the number of times RedMulE accesses the matrices the performance increase again, then drops with parabolic behaviour because the performance is inversely proportional to \textit{P}.

\begin{figure*}[t]
    \centering
    \includegraphics[width=\textwidth]{NetworkPerf3.pdf}
    \vspace{-2mm}
    \caption{RedMulE performance on neural networks compared with software executed on 8 \riscv cores: a) ResNet8 and b) MobileNet V2 training execution.}
    \label{fig:netperf}
\end{figure*}

\subsubsection{FP16 Network Training}
To further evaluate RedMulE performance on a real-case \gls{nn} training, our target is TinyMLPerf~\cite{banbury2021mlperf}, and in particular, we focused on the ResNet~\cite{resnet} example. For the software infrastructure, we rely on the pulp-TrainLib~\cite{trainlib}, and we compared RedMulE with a software baseline executed on 8 \riscv cores sharing 4 \glspl{fpu}. The library takes into consideration all the training steps for the calculation of the gradients and backpropagation. Fig.~\ref{fig:netperf}a shows the execution of a single step in the ResNet8 network when using 8 \riscv cores in parallel and when using RedMulE for the matrix multiplication execution. RedMulE accelerates the matrix multiplication execution of $14.6\times$ with respect to the parallel \riscv execution in SW, speeding up the entire single step of the ResNet8 of $3.1\times$. RedMulE keeps its utilization constant at $99.1\%$ (\SI{47.6}{MAC/cycle}) with the only exceptions in the first and the last layers where it drops to $93.2\%$ (\SI{44.7}{MAC/cycle}) and $32.3\%$ (\SI{15.5}{MAC/cycle}) due to leftovers that do not allow to exploit the full potential of the array. From Fig.~\ref{fig:netperf}a, it is also evident that the data reorganization during the Im2Col accounts for approximately 3 Millions computing cycles. To solve this problem, we augment RedMulE's operation with the support of the DataMover engine, halving the number of computing cycles required to perform the two Im2Col operations and thus speeding up the overall training step execution up to $4.9\times$. As all the devices included in the \gls{pulp} cluster (\riscv cores and accelerators) are designed for synergistic cooperation and share the memory, the heterogeneity of the architecture can be efficiently and fully exploited.

\begin{figure}[t]
    \centering
    \includegraphics[width=\linewidth]{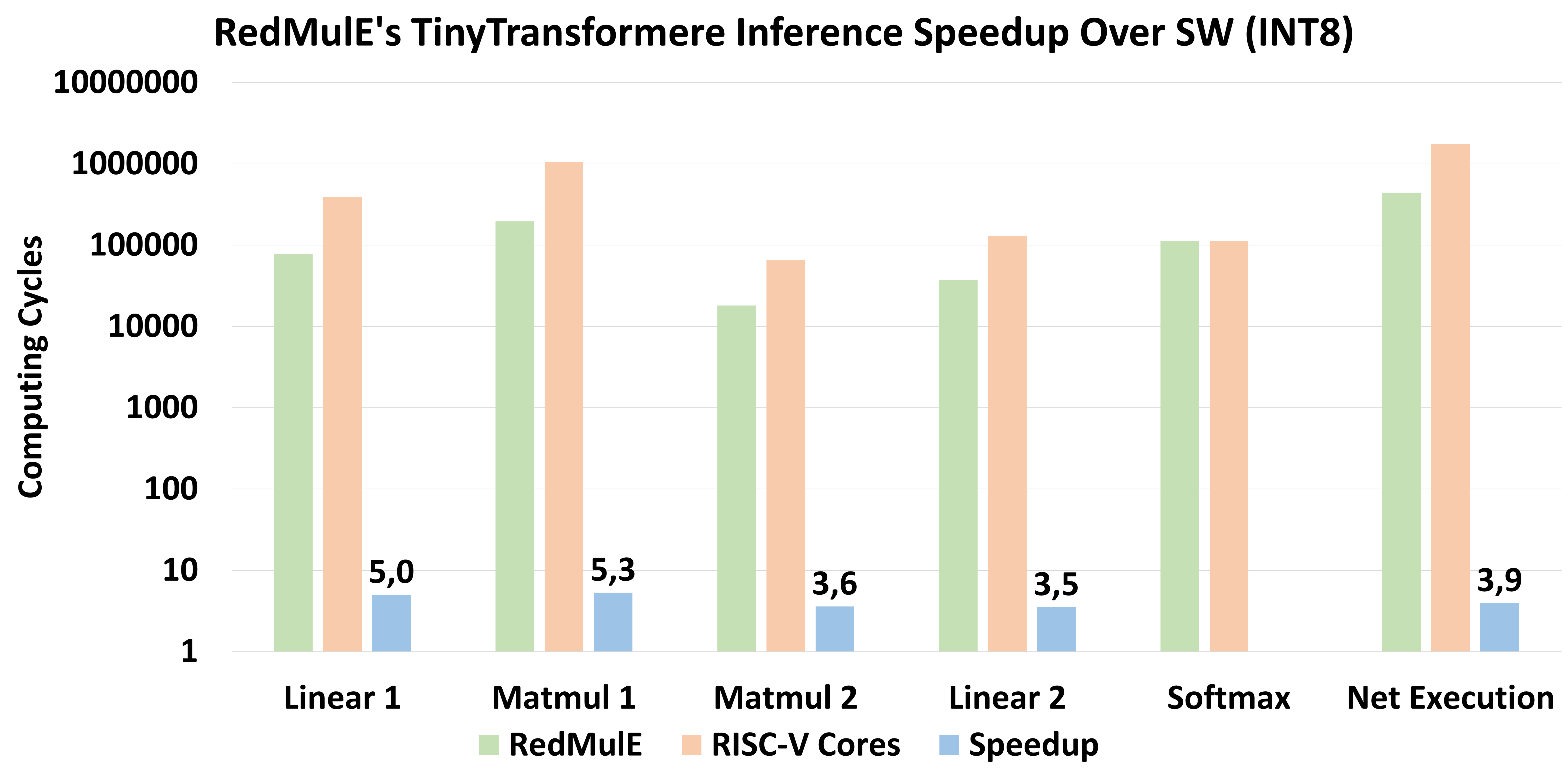}
    \vspace{-2mm}
    \caption{RedMulE performance on Tiny Transformer \gls{fp}8 inference compared with parallel INT8 software executed on 8 \riscv cores.}
    \label{fig:tinytransf}
\end{figure}

\subsubsection{HFP8 Network Training}\label{section:HFP8_train}
For the same training example, we consider a RedMulE\textsubscript{12x8} instance used to train the ResNet network encoded on \dbit{8} \gls{fp} inputs only. For the RedMulE\textsubscript{12x4} we considered until now, the memory port of the \textit{Streamer} is \dbit{288} wide, meaning a \dbit{256} memory port with non-word aligned memory accesses capability. In this configuration, RedMulE\textsubscript{12x4} can load $16\times\gls{fp}16$ elements at a time that are used to fill the pipeline during the computation. Having $H=4$ columns, the pipeline stages within each row are calculated as $H\times(P+1)$, where $P=3$ in this implementation, resulting in 16 pipeline stages. Considering a fixed \dbit{8} input encoding, with the same \dbit{288} memory port, RedMulE can access up to $32\times\gls{fp}8$ elements at a time, meaning that we can implement a RedMule\textsubscript{12x8} instance maintaining a \dbit{288} memory interface and obtaining 32 pipeline stages. We show how the ResNet8 training can benefit from this configuration in the green bar of Fig.~\ref{fig:netperf}a. Matrix multiplication execution can be accelerated up to $28.5\times$, resulting in $5.5\times$ speed-up over the entire training step execution, with $97\%$ utilization.

In addition to the ResNet example, Fig.~\ref{fig:netperf}b shows our evaluation of the speedup introduced by RedMulE over parallel software during the training step of a MobileNet V2~\cite{mobilenetv2} using \gls{fp}8 format. RedMulE introduces a $7.5\times$ average speedup with an $11.2\times$ peak speedup over a carefully optimized software execution on 8 \riscv cores with private \glspl{fpu}. Interestingly, RedMulE accelerates the execution of depthwise layers by up to $2.6\times$ despite the unfavourable sizes of the reshaped matrices that result in high underutilization of the accelerator. RedMulE accelerates the whole training step by $6.4\times$ compared to the parallel execution on general-purpose cores.

\begin{figure}[t]
    \centering
    \includegraphics[width=0.9\linewidth]{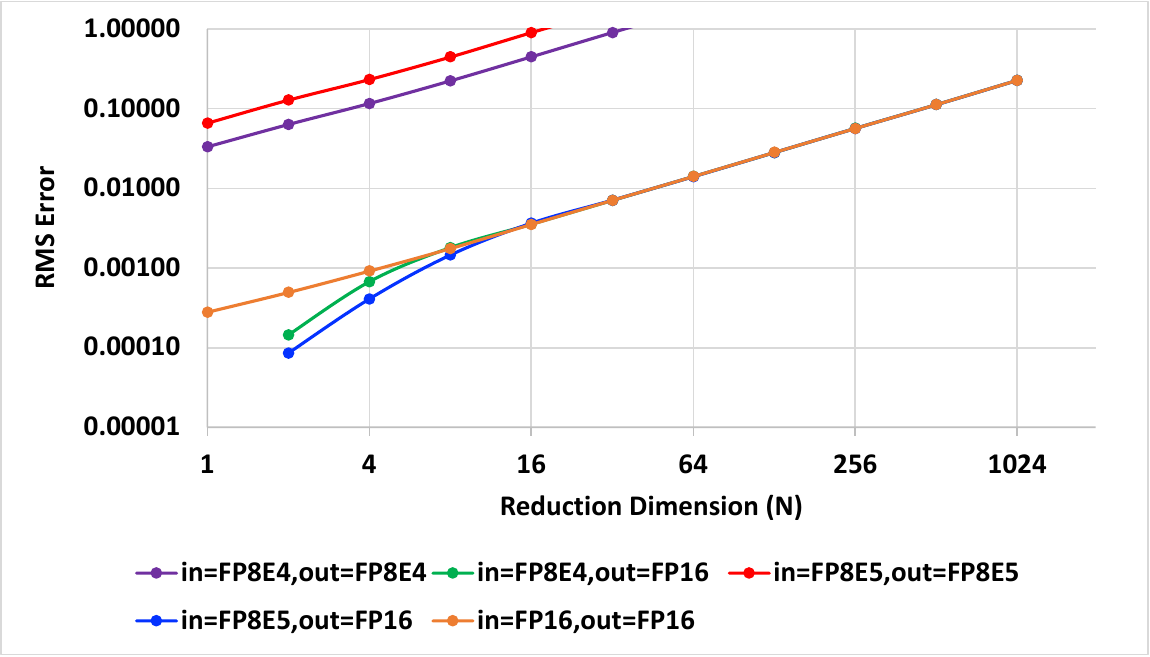}
    \vspace{-2mm}
    \caption{RMS Error analysis for different input/output floating point formats over variable reduction \textit{N} sizes.}
    \label{fig:rmse}
\end{figure}

\begin{figure*}[t]
    \centering
    \includegraphics[width=\textwidth]{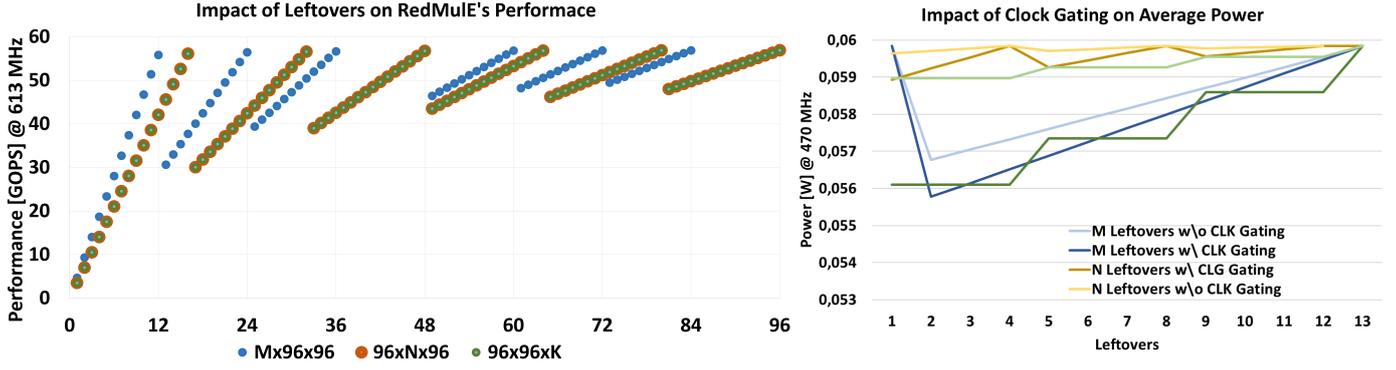}
    \vspace{-2mm}
    \caption{Impact of leftovers on RedMulE performance and energy efficiency with fine-grained clock gating.}
    \label{fig:dynclkgate}
    \vspace{-6mm}
\end{figure*}

\subsubsection{FP8 Transformer Inference}
We evaluated the performance of RedMulE during the execution of a TinyTransformer~\cite{an_mcu} network inference using \gls{fp}8 since it is more suited for transformers as a data format~\cite{fp8train}. Fig.~\ref{fig:tinytransf}  shows the results of the inference of the network, highlighting that the execution of both linear and matmul layers significantly benefits the presence of RedMulE when compared with parallel execution on the general-purpose processors of the cluster. Despite the execution on the general-purpose cores benefits of SIMD extensions with \dbit{8} integer arithmetic, RedMulE introduces more than $4\times$  average speedup on all the network layers, with a peak $5.3\times$ speedup on the Matmul1 layer and a $3.9\times$ speedup on the execution of the entire network.

\begin{figure}[t]
    \centering
    \includegraphics[width=\linewidth]{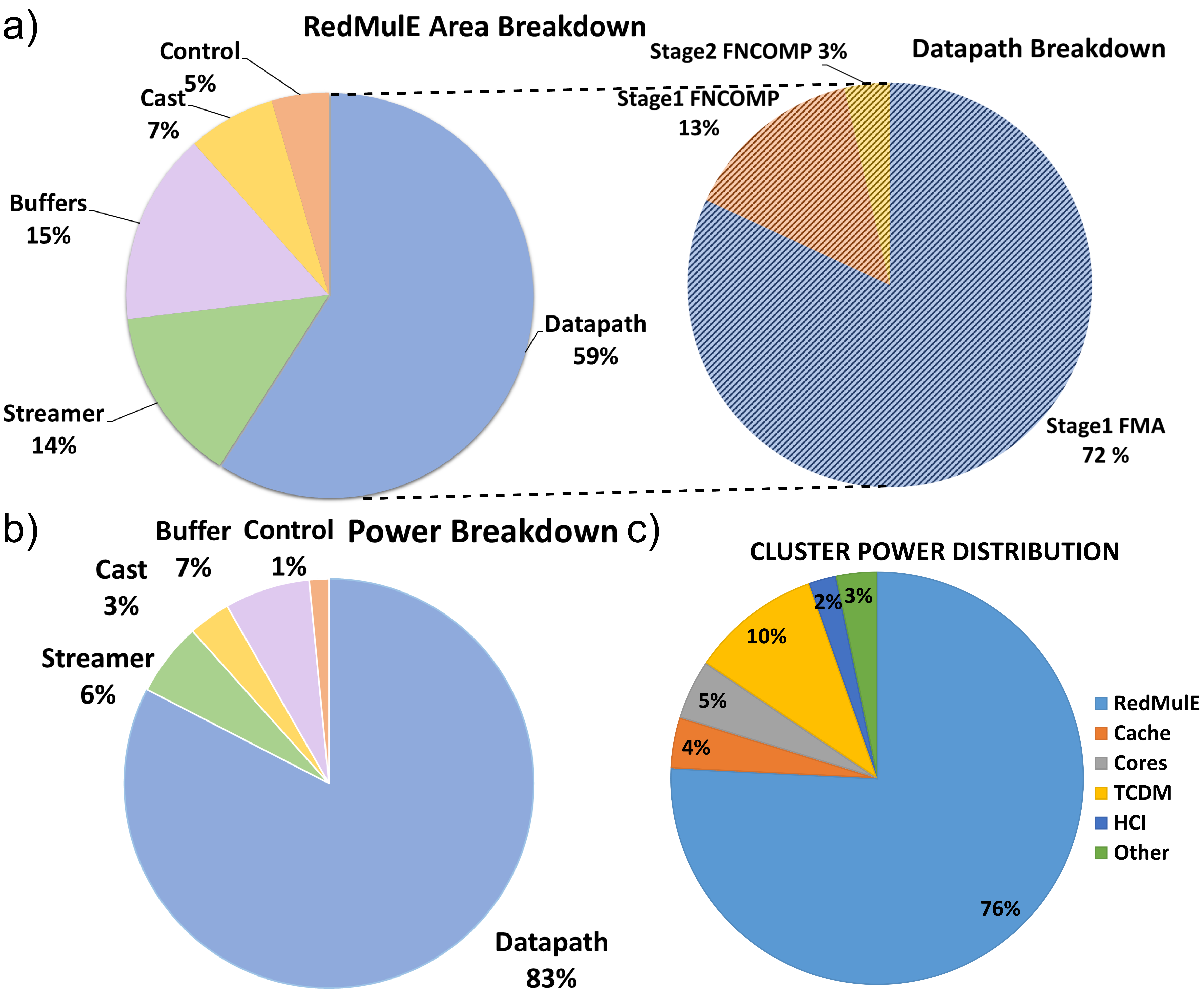}
    \vspace{-2mm}
    \caption{a) RedMulE area breakdown with a focus on the datapath, b) RedMulE power distribution, c) \gls{pulp} power distribution.}
    \label{fig:areapower}
    \vspace{-4mm}
\end{figure}

\subsection{Error Analysis on Input and Output Tensors Format}
We performed error analysis considering an increasing value of the $\mathbf{X}$ and $\mathbf{W}$ matrices' reduction dimension (\textit{N}). The \gls{fma} unit in the \gls{ce} works with fixed internal precision, reducing precision loss during the intermediate accumulation steps. However, when $\mathbf{X}$, $\mathbf{W}$, and $\mathbf{Z}$ are all represented with \dbit{8} precision, the RMSE increases more than $100\times$ compared to the \dbit{16} into \dbit{16} case, with a trend exponentially dependent on the \textit{N} size. However, by representing only $\mathbf{X}$ and $\mathbf{W}$ with \dbit{8} precision and keeping a wider \dbit{16} precision for the output $\mathbf{Z}$-matrix, the accuracy loss is negligible compared to the \dbit{16} only case.

\begin{figure}[t]
    \centering
    \includegraphics[width=\linewidth]{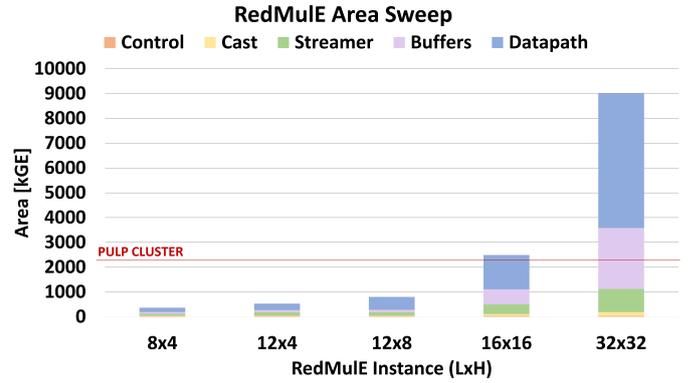}
    \vspace{-2mm}
    \caption{RedMulE area sweep with several sizes of $H$ and $L$.}
    \label{fig:area_sweep}
    \vspace{-4mm}
\end{figure}

\subsection{RedMulE Area}
\subsubsection{Area Breakdown analysis}
RedMulE\textsubscript{12x4} occupies \SI{0.15}{mm^2}, corresponding to $23.8$\% of the entire PULP cluster area (\SI{0.64}{mm^2}). Fig.~\ref{fig:areapower}a shows the breakdown of the RedMulE area, where the cast units account for the $7\%$ to the overall accelerator area, and the \glspl{ce} account for the $59\%$, the $72\%$ of which is given by the \gls{fma} units.

\begin{figure*}[t]
    \centering
    \includegraphics[width=\textwidth]{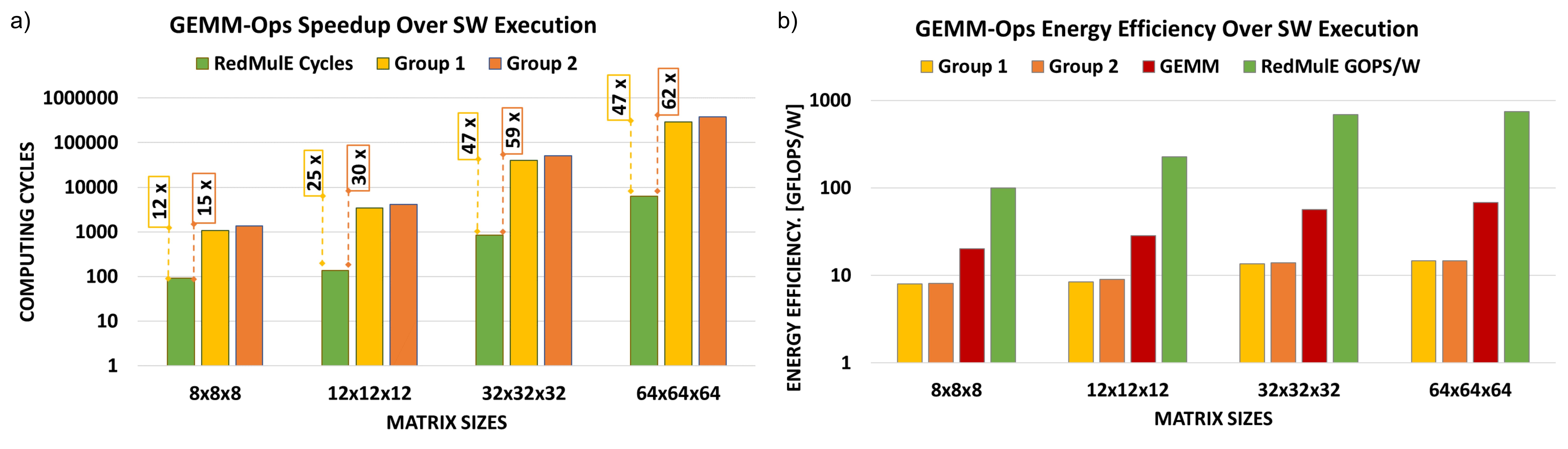}
    \vspace{-2mm}
    \caption{a) Performance and b) energy efficiency of RedMulE during \glspl{gemmop} execution over the parallel software counterpart.}
    \label{fig:gemmops-meas}
    \vspace{-6mm}
\end{figure*}

\subsubsection{RedMulE Area Sweep}
We studied the area overhead introduced when changing the number of \glspl{ce} within RedMulE, fixing the \glspl{ce}' internal pipeline stages to $P=3$. Fig.~\ref{fig:area_sweep} shows that RedMulE's area occupation becomes comparable to the area of the entire \gls{pulp} cluster when it contains 256 \glspl{ce}, corresponding to a RedMulE\textsubscript{16x16} instance. On the other hand, the area of RedMulE\textsubscript{32x32} is $4\times$ larger than the entire \gls{pulp} cluster. Fig.~\ref{fig:area_sweep} shows that changing the shape of the \textit{Datapath} also affects the size of the \textit{Streamer}. In particular, for each \gls{ce} that is added to a row of the \textit{Datapath} (or equivalently, changing the \textit{H} parameter), other $P+1$ pipeline registers are added within each \textit{Datapath} row. The consequence is that the number of elements needed to keep a high \glspl{ce} utilization increases by $P+1$ as well. Keeping $P=3$ as an example, increasing the \textit{H} parameter by 1 requires the \textit{Streamer} to provide $P+1 (= 4)$ additional \gls{fp}16 elements to the \textit{Datapath}. The consequence is that the streamer port must be enlarged of \dbit{64} ($= 4\times{\dbit{16}}$), limiting the integration of RedMulE in the \gls{pulp} cluster.

\subsection{RedMulE Power}
At a cluster level, the power consumption in the efficiency point amounts to \SI{59.3}{mW} during \gls{gemm} operation. The RedMulE contribution dominates the power envelope accounting for $66.8\%$ of the overall consumption, while the \gls{tcdm} banks and the \gls{hci} interconnect contribution is $13.3\%$. In this operating point, we reach a cluster peak energy efficiency of \SI{755}{GFLOPS/W} during \gls{gemm} execution, corresponding to $12.5\times$ higher energy efficiency with respect to the software baseline.
Fig.~\ref{fig:areapower}b and Fig.~\ref{fig:areapower}c show respectively the power breakdown for RedMulE, where most of the power is consumed by the \textit{Datapath}, and the \gls{pulp} cluster during a \gls{gemm} operation, where the majority of the power is consumed by RedMulE and by the \gls{tcdm} banks.

\subsection{Clock Gating Evaluation}
We evaluated how the effect of leftovers impacts the performance and energy efficiency of RedMulE and how fine-grained clock gating helps reduce power consumption during under-utilization phases. Fig.~\ref{fig:dynclkgate} shows our analysis results: to evaluate performance loss, we vary one of the dimensions of the input matrices (\textit{M}, \textit{N}, or \textit{K}) at a time and keep steady the other two, fixing RedMulE's parameters to \textit{L}=12, \textit{H}=4, and \textit{P}=3. To evaluate the power effect, we analyze the power consumption of RedMulE with and without the clock gating contribution in a window corresponding to the last performance step.

In Fig.~\ref{fig:dynclkgate}, the blue dots in the performance diagram highlight the performance trend when we vary the number of $\mathbf{X}$-matrix rows (\textit{M}). Having leftovers in the rows of the $\mathbf{X}$-matrix means that RedMulE uses only a certain number of rows (from one to \textit{L}) to perform the computation of that piece of the matrix. Therefore, an entire part of the RedMulE array does not perform any calculation. In addition, the activity of the \textit{X} and \textit{Z} buffers is also lower because RedMulE needs only part of those buffers. From a performance viewpoint, Fig.~\ref{fig:dynclkgate} shows that the performance rises from \SI{4.7}{GOPS} to \SI{55.8}{GOPS} when \textit{M} goes from 1 to 12. In this case, we have one single tile of the $\mathbf{X}$-matrix being multiplied by all $\mathbf{W}$-matrix tiles. Thus the leftovers impact the entire matrix-matrix operation. By increasing \textit{M}, the number of computing cycles with full array utilization gradually increases, thus balancing the performance losses caused by the presence of leftovers. 

The right part of Fig.~\ref{fig:dynclkgate} shows the power trends resulting from clock gating the unused computing elements in the array and the unused portions of the \textit{X} and \textit{Z} buffers during the leftovers' calculation. The proposed clock gating scheme saves up to $22\%$ of the accelerator's power when the number of $\mathbf{X}$-matrix rows (\textit{M}) is much lower than the number of $\mathbf{X}$ columns (\textit{N}). In this case, a large portion of the \glspl{ce} array is underutilized, and thus clock-gated during most of the computation. Such a situation occurs, for example, during the execution of bottleneck depthwise layers of the MobileNet, which can be reshaped into vector-matrix multiplications ($M=1$, $N >> 1$).

The effect of varying \textit{N} and \textit{K} produces similar results. Leftovers on \textit{N} and \textit{K} turn into a low utilization of the pipeline registers of each accelerator row. From a performance viewpoint, this is similar to the case in which \textit{M} changes. Still, the curves have a lower slope because, with \textit{H}=4 and \textit{P}=3, each row has 16 ($=H\times (P+1)$) internal pipeline registers and every time the value of \textit{N} or \textit{K} reaches a multiple of $H\times (P+1)$ each plot line approaches the maximum performance value. Similarly to the case with variable \textit{M}, also in this case, the first curve starts from a minimum performance value of \SI{3.5}{GOPS} because the leftovers impact the entire computation. From the second to the last lines, the computing cycles in which RedMulE fully uses the array of \glspl{ce} compensate for the impact of matrix leftovers. From the power consumption perspective, the plot lines step once every four matrix elements because RedMulE features four ($=P+1$) pipeline registers between each column of \glspl{ce}. The proposed clock gating methodology allows gating columns of \glspl{ce} that are not active during leftovers' computation so that there can be from one (the first step in the power consumption lines) to four (the last step in each power consumption line) \glspl{ce} active when calculating leftovers. The proposed clock gating scheme saves up to $37\%$ of average power in heavy underutilization conditions, thus increasing the power efficiency by $60\%$ compared to the non-clock-gated case.

\subsection{GEMM-Ops Measurements}
To evaluate the \glspl{gemmop} performance, in Fig.~\ref{fig:gemmops-meas}a we compare the RedMulE \glspl{gemmop} execution against parallel SW execution on the \riscv cores. RedMulE always takes the same number of computing cycles to perform any of the supported \glspl{gemmop}. On the contrary, the parallel execution on the general-purpose cores changes depending on the executed kernel. All the kernels belonging to Group 1 (see Table~\ref{tab:gemmops-table}), i.e. for which the $\circ$ operation is +/x and the $\star$ one is max/min, require the same number of computing cycles when executed on the cores. The execution of these kernels on RedMulE allows for up to $47\times$ speedup. When also $\star$ are of the max/min kind, i.e. Group 2, the execution overhead for the general-purpose cores is even higher, and RedMulE can accelerate such kernels up to $62\times$.

Fig.~\ref{fig:areapower}a shows that the support for \glspl{gemmop} introduces an area overhead of just $16\%$ over the entire accelerator area. $13\%$ of this overhead resides in the first stage \gls{fncomp} unit, dominated by the pipeline introduced to match the latency cycles of the \gls{fma} unit. The second stage \gls{fncomp} unit is fully combinational and accounts only for $3\%$ overhead.

We evaluated the power consumption and the energy efficiency of RedMulE during  \glspl{gemmop} execution. Fig.~\ref{fig:gemmops-meas}b compares the energy efficiency of RedMulE with the software baseline executed on 8 \riscv cores with 4 shared \glspl{fpu} on \gls{fp}16 elements during the execution of \gls{gemm}, \glspl{gemmop}' Group 1 and \glspl{gemmop}' Group 2 kernels. For the \glspl{gemmop}' Group 1, the cluster-level power dissipation reaches \SI{53.2}{mW}, leading to \SI{842}{GFLOPS/W}, $57.2\times$ better than the parallel SW execution on general-purpose cores. On the other hand, during the execution of the algorithms in \glspl{gemmop}' Group 2, the power consumption is further reduced to \SI{37.6}{mW} resulting in \SI{1.19}{TFLOPS/W}, with an energy efficiency increase of $81.2\times$ compared to the parallel software counterpart.

\section{Comparison with the State-of-the-Art}\label{section:soa}

\begin{figure}[t]
    \centering
    \includegraphics[width=\linewidth]{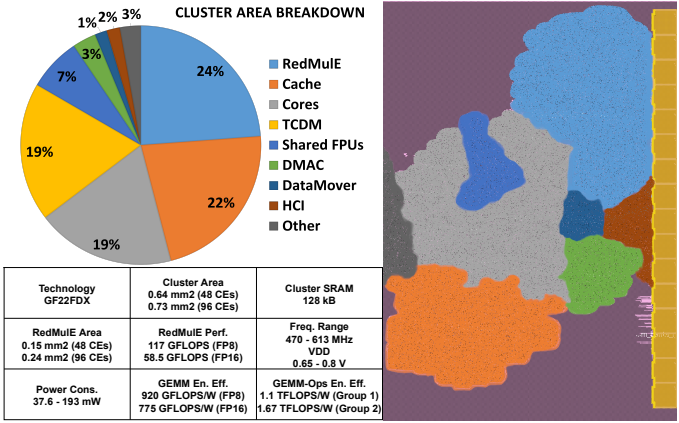}
    \vspace{-2mm}
    \caption{Area breakdown of the \gls{pulp} cluster, layout and resume table.}
    \label{fig:results}
    \vspace{-4mm}
\end{figure}

Table~\ref{table:comparison} resumes the comparison of our work with different \gls{soa} architectures.

We compare our work with \gls{gpu} architectures, in particular with NVIDIA H100 containing TensorCores, that guarantee up to \SI{989}{TFLOPS} of performance in \gls{fp}16 and \SI{1979}{TFLOPS} in \gls{fp}8, meaning $17000\times$ than our work, but at the cost of \SI{700}{W} power consumption and \SI{814}{mm^2}, $12000\times$ more power-consuming and $1300\times$ larger than our work -- representing an unfeasible solution for an \gls{iot} end-node.

While RedMulE targets primarily training, it is also usable for inference. For this reason, we include in our comparison some inference-oriented chips, like DNPU~\cite{dnpu}. DNPU's performance is just $1.9\times$ higher than our cluster, although DNPU contains $16\times$ the number of \glspl{ce}. Moreover, DNPU features $2.7\times$ higher efficiency than RedMulE but is designed to work with fixed-point precision only, which helps increase energy efficiency. We also compared our work with Diana~\cite{diana} and Gemmini~\cite{gemmini}, being designed in the same technology node of RedMulE. The former achieves $44.5\%$ less performance than \redmuleByte and $12\%$ less performance than \redmuleHalf in the energy efficient mode. Diana's power consumption in efficiency mode is much lower than our design, but if we scale down the frequency to 50 MHz as they do, our \gls{pulp} cluster with \redmuleHalf consumes just \SI{7.65}{mW}. The significant increase in Diana's energy efficiency is justified as it uses only \dbit{8} integer arithmetic, which helps reduce power consumption and increases energy efficiency. On the contrary, Gemmini features one order of magnitude less energy efficiency than \redmuleHalf despite it features $5\times$ the number of \glspl{ce} and works with \dbit{8} integer format.

We also compared our design with other platforms specifically designed for on-chip training. IBM~\cite{4c_ibm} demonstrated a 4-core AI chip in \SI{7}{nm} technology which is just $2.4\times$ more energy-efficient, $33.2\times$ larger, and $74\times$ more power-consuming than our PULP cluster with \redmuleHalf, despite the technology scaling and the reduced operating voltage. IBM also proposes a chip~\cite{oh_ibm}, with more than \SI{1}{W} of power consumption, which is not acceptable for extreme-edge training. On the other hand, LNPU~\cite{lnpu} is an extreme-edge processor that features a $6.67\times$ higher power envelope than \redmuleHalf. Vega is a valid candidate for on-chip embedded training, but \redmuleHalf achieves $7.8\times$ higher performance and $3.2\times$ higher energy efficiency. Cambricon-Q~\cite{cambricon} is designed in \SI{45}{nm} and is $2.9\times$ more energy-efficient than our design but makes use of narrow \dbit{8} fixed-point arithmetic, while generally available learning algorithms based on backpropagation strictly require \gls{fp} range and precision. Cambricon-Q is also $17.7\times$ more power-hungry than our design, therefore not suitable for Tiny\gls{ml} applications. Similar considerations hold for T-PIM~\cite{tpim}, a training chip designed in \SI{28}{nm} technology that features an in-memory computing core for high energy efficiency but only works with \dbit{16} integer precision, not satisfying the precision requirements to enable on-chip training.

\begin{table*}
\centering
\caption{State of the art comparison. First line = Best Efficiency; Second line = Peak Performance. 1 MAC = 2 OPs.}
\label{table:comparison}
\begin{tabular}{|c|c|c|c|c|c|c|c|c|c|c|}
\hline
\textbf{Category} &
  \textbf{Design} &
  \textbf{\begin{tabular}[c]{@{}c@{}}Tech\\ \textit{nm}\end{tabular}} &
  \textbf{\begin{tabular}[c]{@{}c@{}}Area\\ \textit{mm${}{^2}$}\end{tabular}} &
  \textbf{\begin{tabular}[c]{@{}c@{}}Freq\\ \textit{MHz}\end{tabular}} &
  \textbf{\begin{tabular}[c]{@{}c@{}}Volt\\ \textit{V}\end{tabular}} &
  \textbf{\begin{tabular}[c]{@{}c@{}}Power\\ \textit{mW}\end{tabular}} &
  \textbf{\begin{tabular}[c]{@{}c@{}}Perf\\ \textit{GOPS}\end{tabular}} &
  \textbf{\begin{tabular}[c]{@{}c@{}}Energy Eff\\ \textit{GOPS/W}\end{tabular}} &
  \textbf{CEs} &
  \textbf{Precision} \\ \hline
\multirow{2}{*}{GPU} &
  NVIDIA H100~\cite{h_100} &
  4 &
  814 &
  1830 &
  - &
  700000 &
  \begin{tabular}[c]{@{}c@{}}1978900\\ 989400\end{tabular} &
  \begin{tabular}[c]{@{}c@{}}2827\\ 1413\end{tabular} &
  528 &
  \begin{tabular}[c]{@{}c@{}}FP8\\ FP16\end{tabular} \\ \cline{2-11} 
 &
  SIMD${}{^2}$~\cite{simd2} &
  45 &
  19.5 &
  - &
  - &
  4190 &
  - &
  - &
  - &
  INT16 \\ \hline
\multirow{3}{*}{\begin{tabular}[c]{@{}c@{}}Inference\\ Chips\end{tabular}} &
  DNPU~\cite{dnpu} &
  65 &
  16 &
  \begin{tabular}[c]{@{}c@{}}50\\ 200\end{tabular} &
  \begin{tabular}[c]{@{}c@{}}0.7\\ 1.1\end{tabular} &
  \begin{tabular}[c]{@{}c@{}}34.6\\ 279\end{tabular} &
  \begin{tabular}[c]{@{}c@{}}72.6\\ 279\end{tabular} &
  \begin{tabular}[c]{@{}c@{}}2100\\ 1000\end{tabular} &
  768 &
  INT16 \\ \cline{2-11} 
 &
  Diana~\cite{diana} &
  22 &
  8.91 &
  \begin{tabular}[c]{@{}c@{}}50\\ 280\end{tabular} &
  \begin{tabular}[c]{@{}c@{}}0.55\\ 0.9\end{tabular} &
  \begin{tabular}[c]{@{}c@{}}9.96\\ 129\end{tabular} &
  \begin{tabular}[c]{@{}c@{}}40\\ 224\end{tabular} &
  \begin{tabular}[c]{@{}c@{}}4040\\ 1740\end{tabular} &
  256 &
  INT8 \\ \cline{2-11} 
 &
  Gemmini~\cite{gemmini} &
  22 &
  16 &
  \begin{tabular}[c]{@{}c@{}}700\\ 900\end{tabular} &
  \begin{tabular}[c]{@{}c@{}}0.75\\ 0.91\end{tabular} &
  \begin{tabular}[c]{@{}c@{}}-\\ -\end{tabular} &
  \begin{tabular}[c]{@{}c@{}}-\\ -\end{tabular} &
  \begin{tabular}[c]{@{}c@{}}70\\ 50\end{tabular} &
  256 &
  INT8 \\ \hline
\multirow{11}{*}{\begin{tabular}[c]{@{}c@{}}Training\\ Chips\end{tabular}} &
  4-core IBM~\cite{4c_ibm} &
  7 &
  19.6 &
  \begin{tabular}[c]{@{}c@{}}1000\\ 1600\end{tabular} &
  \begin{tabular}[c]{@{}c@{}}0.55\\ 0.75\end{tabular} &
  \begin{tabular}[c]{@{}c@{}}4400\\ 13000\end{tabular} &
  \begin{tabular}[c]{@{}c@{}}8000\\ 12800\end{tabular} &
  \begin{tabular}[c]{@{}c@{}}1800\\ 980\end{tabular} &
  4096 &
  FP16 \\ \cline{2-11} 
 &
  LNPU~\cite{lnpu} &
  65 &
  16 &
  200 &
  \begin{tabular}[c]{@{}c@{}}0.78\\ 1.1\end{tabular} &
  367 &
  \begin{tabular}[c]{@{}c@{}}600\\ 300\end{tabular} &
  \begin{tabular}[c]{@{}c@{}}1630\\ 817\end{tabular} &
  768 &
  \begin{tabular}[c]{@{}c@{}}FP8\\ FP16\end{tabular} \\ \cline{2-11} 
 &
  Oh, IBM~\cite{oh_ibm} &
  14 &
  9.8 &
  \begin{tabular}[c]{@{}c@{}}1000\\ 1500\end{tabular} &
  \begin{tabular}[c]{@{}c@{}}0.54\\ 0.62\end{tabular} &
  \begin{tabular}[c]{@{}c@{}}1428\\ 2727\end{tabular} &
  \begin{tabular}[c]{@{}c@{}}2000\\ 3000\end{tabular} &
  \begin{tabular}[c]{@{}c@{}}1400\\ 1100\end{tabular} &
  128 &
  \begin{tabular}[c]{@{}c@{}}FP32\\ FP16\end{tabular} \\ \cline{2-11} 
 &
  T-PIM~\cite{tpim} &
  28 &
  5.04 &
  \begin{tabular}[c]{@{}c@{}}50\\ 280\end{tabular} &
  \begin{tabular}[c]{@{}c@{}}0.75\\ 1.05\end{tabular} &
  \begin{tabular}[c]{@{}c@{}}5.25\\ 51.2\end{tabular} &
  \begin{tabular}[c]{@{}c@{}}39.8\\ 43\end{tabular} &
  \begin{tabular}[c]{@{}c@{}}7590\\ 840\end{tabular} &
  - &
  INT16 \\ \cline{2-11} 
 &
  \multirow{2}{*}{TSUNAMI~\cite{tsunami}} &
  \multirow{2}{*}{65} &
  \multirow{2}{*}{16} &
  \multirow{2}{*}{\begin{tabular}[c]{@{}c@{}}50\\ 200\end{tabular}} &
  \multirow{2}{*}{\begin{tabular}[c]{@{}c@{}}0.78\\ 1.1\end{tabular}} &
  \multirow{2}{*}{\begin{tabular}[c]{@{}c@{}}45\\ 419\end{tabular}} &
  612 &
  \begin{tabular}[c]{@{}c@{}}3420\\ 1480\end{tabular} &
  2048 &
  FP8 \\ \cline{8-11} 
 &
   &
   &
   &
   &
   &
   &
  306 &
  \begin{tabular}[c]{@{}c@{}}1710\\ 740\end{tabular} &
  1024 &
  FP16 \\ \cline{2-11} 
 &
  \multirow{2}{*}{Trainer~\cite{trainer}} &
  \multirow{2}{*}{28} &
  \multirow{2}{*}{21} &
  \multirow{2}{*}{\begin{tabular}[c]{@{}c@{}}40\\ 440\end{tabular}} &
  \multirow{2}{*}{\begin{tabular}[c]{@{}c@{}}0.56\\ 1\end{tabular}} &
  \multirow{2}{*}{\begin{tabular}[c]{@{}c@{}}23\\ 363\end{tabular}} &
  900 &
  4280 &
  8192 &
  FP8 \\ \cline{8-11} 
 &
   &
   &
   &
   &
   &
   &
  450 &
  2140 &
  4096 &
  FP16 \\ \cline{2-11} 
 &
  Cambricon-Q~\cite{cambricon} &
  45 &
  888 &
  1000 &
  0.6 &
  1030 &
  2000 &
  2240 &
  1024 &
  INT8 \\ \cline{2-11} 
 &
  \multirow{2}{*}{Vega~\cite{vega}} &
  \multirow{2}{*}{22} &
  \multirow{2}{*}{12} &
  \multirow{2}{*}{450} &
  \multirow{2}{*}{\begin{tabular}[c]{@{}c@{}}0.5\\ 0.8\end{tabular}} &
  \multirow{2}{*}{\begin{tabular}[c]{@{}c@{}}-\\ 49.4\end{tabular}} &
  \multirow{2}{*}{\begin{tabular}[c]{@{}c@{}}3.3\\ 7.5\end{tabular}} &
  \multirow{2}{*}{\begin{tabular}[c]{@{}c@{}}250\\ 180\end{tabular}} &
  \multirow{2}{*}{4} &
  \multirow{2}{*}{FP16} \\
 &
   &
   &
   &
   &
   &
   &
   &
   &
   &
   \\ \hline
Mat-Mul &
  Anders~\cite{anders} &
  14 &
  0.024 &
  \begin{tabular}[c]{@{}c@{}}2.1\\ 1090\end{tabular} &
  \begin{tabular}[c]{@{}c@{}}0.26\\ 0.9\end{tabular} &
  \begin{tabular}[c]{@{}c@{}}0.023\\ 82.7\end{tabular} &
  \begin{tabular}[c]{@{}c@{}}0.068\\ 34\end{tabular} &
  \begin{tabular}[c]{@{}c@{}}2970\\ 420\end{tabular} &
  16 &
  FP16 \\ \hline
\textbf{GEMM} &
  \multirow{3}{*}{\textbf{\begin{tabular}[c]{@{}c@{}}This Work\\ RedMulE\textsubscript{12x4}\end{tabular}}} &
  \multirow{3}{*}{\textbf{22}} &
  \multirow{3}{*}{\textbf{0.64}} &
  \multirow{3}{*}{\textbf{\begin{tabular}[c]{@{}c@{}}470\\ 613\end{tabular}}} &
  \multirow{3}{*}{\textbf{\begin{tabular}[c]{@{}c@{}}0.65\\ 0.8\end{tabular}}} &
  \textbf{\begin{tabular}[c]{@{}c@{}}59.3\\ 116\end{tabular}} &
  \multirow{3}{*}{\textbf{\begin{tabular}[c]{@{}c@{}}44.8\\ 58.5\end{tabular}}} &
  \textbf{\begin{tabular}[c]{@{}c@{}}775\\ 506\end{tabular}} &
  \multirow{3}{*}{\textbf{48}} &
  \multirow{3}{*}{\textbf{FP16}} \\ \cline{1-1} \cline{7-7} \cline{9-9}
\textbf{\begin{tabular}[c]{@{}c@{}}Group 1\\ GEMM-Ops\end{tabular}} &
   &
   &
   &
   &
   &
  \textbf{\begin{tabular}[c]{@{}c@{}}53.2\\ 103\end{tabular}} &
   &
  \textbf{\begin{tabular}[c]{@{}c@{}}842\\ 576\end{tabular}} &
   &
   \\ \cline{1-1} \cline{7-7} \cline{9-9}
\textbf{\begin{tabular}[c]{@{}c@{}}Group 2\\ GEMM-Ops\end{tabular}} &
   &
   &
   &
   &
   &
  \textbf{\begin{tabular}[c]{@{}c@{}}37.6\\ 71.5\end{tabular}} &
   &
  \textbf{\begin{tabular}[c]{@{}c@{}}1193\\ 819\end{tabular}} &
   &
   \\ \hline
\textbf{GEMM} &
  \multirow{3}{*}{\textbf{\begin{tabular}[c]{@{}c@{}}This Work\\ RedMulE\textsubscript{12x8}\end{tabular}}} &
  \multirow{3}{*}{\textbf{22}} &
  \multirow{3}{*}{\textbf{0.73}} &
  \multirow{3}{*}{\textbf{\begin{tabular}[c]{@{}c@{}}470\\ 613\end{tabular}}} &
  \multirow{3}{*}{\textbf{\begin{tabular}[c]{@{}c@{}}0.65\\ 0.8\end{tabular}}} &
  \textbf{\begin{tabular}[c]{@{}c@{}}97.5\\ 193\end{tabular}} &
  \multirow{3}{*}{\textbf{\begin{tabular}[c]{@{}c@{}}89.7\\ 117\end{tabular}}} &
  \textbf{\begin{tabular}[c]{@{}c@{}}920\\ 608\end{tabular}} &
  \multirow{3}{*}{\textbf{96}} &
  \multirow{3}{*}{\textbf{FP8}} \\ \cline{1-1} \cline{7-7} \cline{9-9}
\textbf{\begin{tabular}[c]{@{}c@{}}Group 1\\ GEMM-Ops\end{tabular}} &
   &
   &
   &
   &
   &
  \textbf{\begin{tabular}[c]{@{}c@{}}85.2\\ 168\end{tabular}} &
   &
  \textbf{\begin{tabular}[c]{@{}c@{}}1052\\ 694\end{tabular}} &
   &
   \\ \cline{1-1} \cline{7-7} \cline{9-9}
\textbf{\begin{tabular}[c]{@{}c@{}}Group 2\\ GEMM-Ops\end{tabular}} &
   &
   &
   &
   &
   &
  \textbf{\begin{tabular}[c]{@{}c@{}}54\\ 104\end{tabular}} &
   &
  \textbf{\begin{tabular}[c]{@{}c@{}}1666\\ 1123\end{tabular}} &
   &
   \\ \hline
\end{tabular}
\end{table*}

TSUNAMI~\cite{tsunami} and Trainer~\cite{trainer} are conceived for energy-efficient embedded training and extensively use pruning and sparse matrices generation to increase energy efficiency and reduce the number of required MAC operations during training with zero-skipping. We compare with the results they provide during dense calculations. 
In their best efficiency points, TSUNAMI and Trainer's power consumption is comparable to RedMulE's. However, those points correspond to \SI{50}{MHz} and \SI{40}{MHz} for TSUNAMI and Trainer, while RedMule is evaluated at \SI{470}{MHz}. Therefore, RedMulE would consume approximately one order of magnitude less power at comparable frequencies.
TSUNAMI and Trainer reach up to $5\times$ and $8\times$ higher performance, respectively, since they feature $21\times$ and $85\times$ the number of \glspl{ce}, but feature a much lower \glspl{ce}' utilization than our RedMulE ($75\%$ TSUNAMI and only $ 12.5\%$ Trainer). The systolic architecture of RedMulE enables, in principle, almost arbitrary architecture scaling. Assuming linear performance, area, and power ratio, scaling to 1024 or 4096 \glspl{ce} ($21\times$ and $85\times$ larger than \redmuleHalf), our utilization would still be $99.4\%$, leading to higher overall performance (\SI{1.25}{TFLOPS} and \SI{5}{TFLOPS} respectively) once accounting overheads.

We compare \redmuleHalf with Anders \etal~\cite{anders}, proposing a hardware accelerator for matrix multiplications in \SI{14}{nm} technology that targets TinyML learning and inference applications. It reaches a peak energy efficiency of \SI{2.97}{TFLOPS/W} in \gls{fp}16 precision, $3.83\times$ higher than \redmuleHalf, but only when operating at near-threshold voltage (\SI{260}{mV}) and extremely reduced frequency (\SI{2.1}{MHz}). In that operating point, their design is $659\times$ less performant than RedMulE. Anders' peak performance is obtained at \SI{0.9}{V} and \SI{1.09}{GHz}, leading to \SI{420}{GFLOPS/W} and $99.4$\% MAC units utilization, similarly to RedMulE's. In \SI{22}{nm} technology, at \SI{613}{MHz} frequency and \SI{0.8}{V}, \redmuleHalf reaches \SI{58.5}{GFLOPS}, $1.72\times$ better than Anders \etal, with a $20.5\%$ higher energy efficiency of \SI{506}{GFLOPS/W} on \gls{fp}16 \gls{gemm}.

We also compared RedMulE with \simdsquare~\cite{simd2}, the only other design that features \glspl{gemmop} extensions. Even though \simdsquare works only with integer arithmetic and cannot thus target on-chip training, it features $36.1\times$ higher power consumption than RedMulE. In their case, the authors also claim that the area overhead to build \glspl{gemmop} extensions on top of NVIDIA Streaming Multiprocessor accounted for $69\%$, while in RedMulE, the area overhead introduced by \glspl{gemmop} extension is just $16\%$.

\section{Conclusion}\label{section:conclusions}

In this paper, we presented RedMulE - Reduced-Precision Matrix Multiplication Engine, a fully-parametric open-source cluster-coupled accelerator enabling TinyML training on ultra-low-power devices, i.e. near-sensor training on a few tens of mW of power budget. RedMulE is
conceived for \gls{fp}16 \glspl{gemmop} computation, and supports compressed \gls{fp}8 inputs while also efficiently accelerating a wide range of operations that share the same structure of a \gls{gemm}. RedMulE allows the instantiation of a wide range of Floating-Point Units-based Computing Elements (CEs), internal buffers, and memory interface configurations. We integrated an instance of RedMulE, containing a $12\times4$  array of \glspl{ce} into an ultra-low-power cluster containing 8 \riscv cores, and implemented the resulting system in a \SI{22}{nm} technology. 
RedMulE achieves $99.4\%$ \glspl{ce} utilization and  an average $15\times$ speedup during simple \gls{gemm} execution with respect to a parallel software baseline running on the eight cores. It occupies \SI{0.15}{mm^2} accounting for only $24$\% of the cluster area. During \glspl{gemmop} execution, the performance speedup introduced by RedMulE over the \riscv cores reaches up to $62\times$.
In its best performance point (at \SI{613}{MHz}, \SI{0.8}{V}), RedMulE achieves \SI{506}{GFLOPS/W} @ \SI{58.5}{GFLOPS} when executing \gls{fp} \gls{gemm} kernels; while, in its best efficiency point (at \SI{470}{MHz}, \SI{0.65}{V}), it reaches \SI{775}{GFLOPS/W} @ \SI{44.8}{GFLOPS}. 
On a real example of \gls{nn} training, RedMulE accelerates the matrix multiplication by up to $14.6\times$ and $28.5\times$ when the input tensors are represented with \dbit{16} and \dbit{8} respectively, accelerating the whole training step of ResNet8 by $4.9\times$ and $5.2\times$.

\section*{Acknowledgments}\label{section:ack}
This work was supported in part by Thales Alenia Space, The European PILOT (EuroHPC JU, g.a. 101034126), and NeuroSoC (Horizon EU g.a. 101070634). Supported in part by “TinyTrainer” project that receives funding from the Swiss National Science Foundation (g.a. 20791).

\bibliography{biblio}

\par\noindent 
\parbox[t]{\linewidth}{
\noindent{\includegraphics[height=1.5in,width=1in,clip,keepaspectratio]{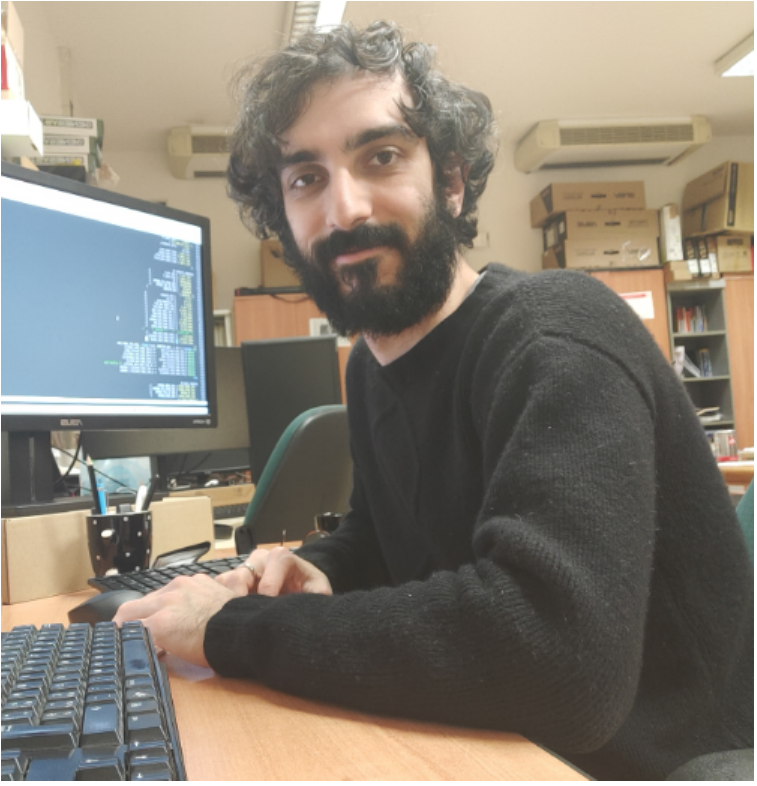}}
\noindent {\bf Yvan Tortorella}\
received his Master's Degree in Electronic Engineering in October 2021 from the University of Bologna. He is currently pursuing a Ph. D. in Digital Systems Design in the group of Professor Luca Benini at the Department of Electrical and Information Engineering (DEI) of the University of Bologna. His research interests include the design of PULP (Parallel Ultra-Low Power)-based hardware accelerators for ultra-low power Machine Learning and the design of RISC-V-based computer architectures for satellite applications.}
\vspace{4\baselineskip}

\par\noindent 
\parbox[t]{\linewidth}{
\noindent{\includegraphics[height=1.5in,width=1in,clip,keepaspectratio]{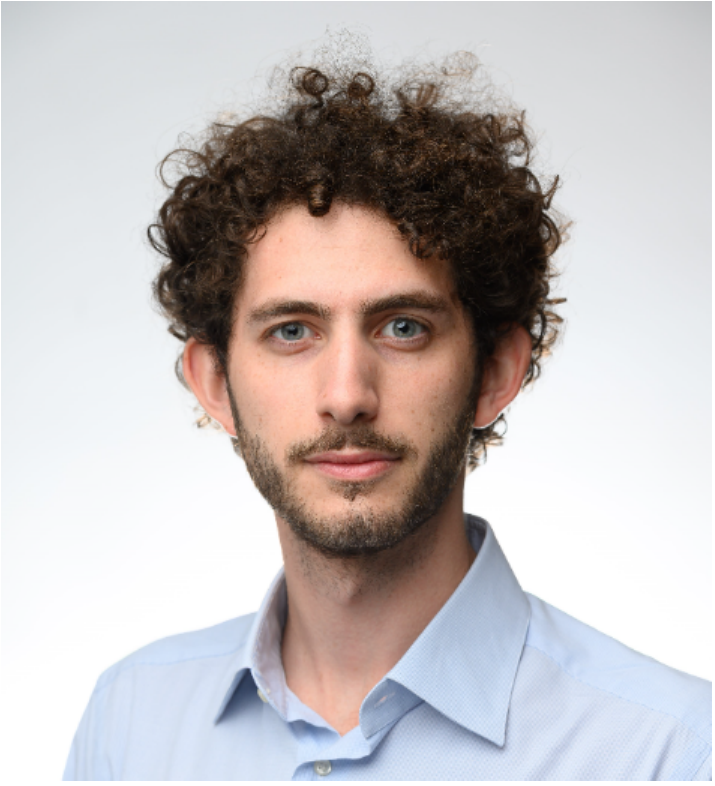}}
\noindent {\bf Luca Bertaccini}\
received the M.Sc. degree in Electronic Engineering from the University of Bologna in 2020. He is currently pursuing a Ph.D. degree at ETH Zürich in the Digital Circuits and Systems group led by Prof. Luca Benini. His research interests include heterogeneous systems-on-chip, energy-efficient hardware accelerators, computer arithmetic, and transprecision computing. He received the 2021 IEEE ASAP Best Paper Honorable Mention.}
\vspace{4\baselineskip}

\par\noindent 
\parbox[t]{\linewidth}{
\noindent{\includegraphics[height=1.5in,width=1in,clip,keepaspectratio]{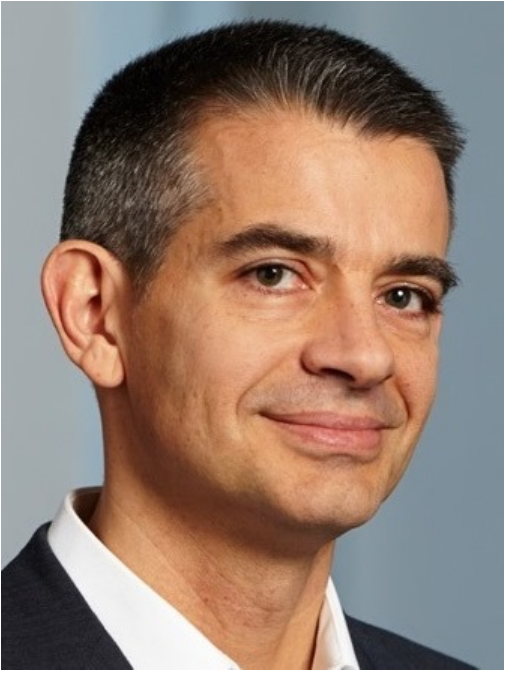}}
\noindent {\bf Luca Benini}\
holds the chair of digital Circuits and systems at ETHZ and is Full Professor at the Università di Bologna. He received a PhD from Stanford University. Dr. Benini's research interests are in energy-efficient parallel computing systems, smart sensing micro-systems and machine learning hardware. He has published more than 1000 peer-reviewed papers and five books. He is a Fellow of the IEEE, of the ACM and a member of the Academia Europaea. He received the IEEE Mac Van Valkenburg award in 2016 and the ACM/IEEE A. Richard Newton Award in 2020.}
\vspace{4\baselineskip}

\par\noindent 
\parbox[t]{\linewidth}{
\noindent{\includegraphics[height=1.5in,width=1in,clip,keepaspectratio]{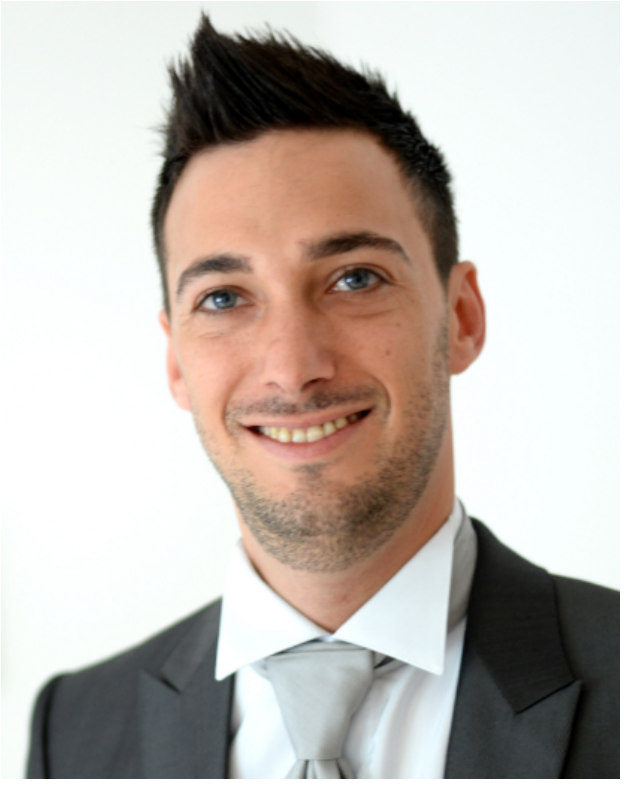}}
\noindent {\bf Davide Rossi}\
received the Ph.D. degree from the University of Bologna, Bologna, Italy, in 2012. He has been a Post-Doctoral Researcher with the Department of Electrical, Electronic and Information Engineering “Guglielmo Marconi,” University of Bologna, since 2015, where he is currently an Associate Professor. His research interests focus on energy-efficient digital architectures. In this field, he has published more than 100 papers in international peer-reviewed conferences and journals. He is recipient of Donald O. Pederson Best Paper Award 2018, 2020 IEEE TCAS Darlington Best Paper Award, 2020 IEEE TVLSI Prize Paper Award.}
\vspace{4\baselineskip}

\par\noindent 
\parbox[t]{\linewidth}{
\noindent{\includegraphics[height=1.5in,width=1in,clip,keepaspectratio]{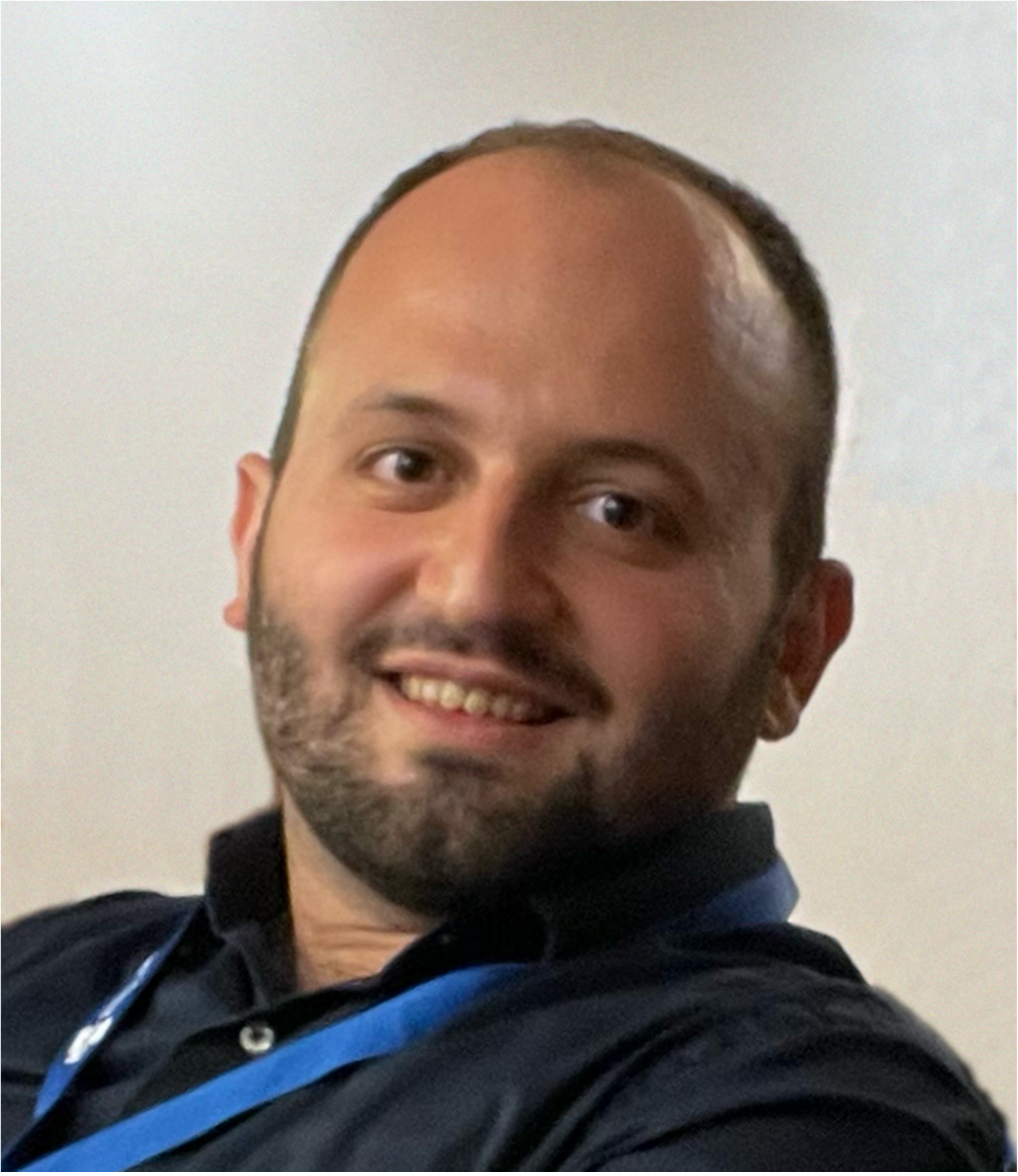}}
\noindent {\bf Francesco Conti}\
received the Ph.D. degree in electronic engineering from the University of Bologna, Italy, in 2016.
He is currently a Tenure-Track Assistant Professor with the DEI Department, University of Bologna. From 2016 to 2020, he held a research grant with the University of Bologna and a Post-Doctoral Researcher with ETH Zürich.
His research is centered on hardware acceleration in ultra-low power and highly energy efficient platforms, with a particular focus on System-on-Chips for Artificial Intelligence applications.
His research work has resulted in more than 70 publications in international conferences and journals and was awarded several times, including the 2020 IEEE TCAS-I Darlington Best Paper Award.}
\vspace{4\baselineskip}

\end{document}